\documentclass[apj]{emulateapj}

\usepackage{graphicx}

\slugcomment{accepted to the Astrophysical Journal 10 August 2011}

\begin{document}

\title{The Major and Minor Galaxy Merger Rates at $z<1.5$}

\author{Jennifer M. Lotz\altaffilmark{1,2,3}, Patrik Jonsson\altaffilmark{4}, 
T.J. Cox\altaffilmark{5,6}, Darren Croton\altaffilmark{7},  
\\
Joel R. Primack\altaffilmark{8},  Rachel S. Somerville\altaffilmark{3, 9}, and Kyle Stewart\altaffilmark{10, 11}}
\altaffiltext{1}{National Optical Astronomical Observatories, 950 N. Cherry Avenue, Tucson, AZ 85719, USA}
\altaffiltext{2}{Leo Goldberg Fellow}
\altaffiltext{3}{Space Telescope Science Institute, 3700 San Martin Dr., Baltimore, MD 21218; lotz@stsci.edu}
\altaffiltext{4}{Harvard-Smithsonian Center for Astrophysics, Cambridge, MA, USA}
\altaffiltext{5}{Carnegie Observatories, Pasadena, CA, USA}
\altaffiltext{6}{Carnegie Fellow}
\altaffiltext{7}{Centre for Astrophysics \& Supercomputing, Swinburne University of Technology,  Hawthorn, Australia}
\altaffiltext{8}{Department of Physics, University of California, Santa Cruz, USA}
\altaffiltext{9}{Department of Physics \& Astronomy, Johns Hopkins University, Baltimore, MD, USA}
\altaffiltext{10}{Jet Propulsion Laboratory, Pasadena, CA, USA}
\altaffiltext{11}{NASA Postdoctoral Fellow}

\begin{abstract}
Calculating the galaxy merger rate requires both a census of galaxies identified as 
merger candidates,  and a cosmologically-averaged `observability' timescale
$\langle T_{obs}(z) \rangle$ for identifying galaxy mergers.   While many have counted 
galaxy mergers using a variety of  techniques, $\langle T_{obs}(z) \rangle$ for these techniques have been poorly constrained.  
We address this problem by calibrating three merger rate estimators with a suite of hydrodynamic merger simulations and 
three galaxy formation models. We estimate $\langle T_{obs}(z) \rangle$ for (1) close galaxy 
pairs with a range of projected separations, (2) the morphology indicator $G-M_{20}$, and (3) the 
morphology indicator  asymmetry $A$.  Then we apply these timescales to the observed 
merger fractions at $z < 1.5$ from the recent literature. When our physically-motivated timescales are adopted, 
the observed galaxy merger rates become largely consistent.   The remaining differences between the galaxy merger rates are explained by
the differences in the range of mass-ratio measured by different techniques and differing parent galaxy selection. 
The major merger rate per unit co-moving volume for samples selected with constant number density evolves much more strongly with
redshift ($\propto (1+z)^{+3.0 \pm 1.1} $) than samples selected with constant stellar mass or passively
evolving luminosity  ($\propto (1+z)^{+0.1 \pm 0.4}$). We calculate the minor merger rate (1:4 $< M_{sat}/M_{primary} \lesssim$ 1:10)
by subtracting the major merger rate from close pairs from the `total' merger rate determined by $G-M_{20}$. 
The implied minor merger rate  is  $\sim 3$ times the major merger rate 
at $z \sim 0.7$, and shows little evolution with redshift.
\end{abstract}


\section{INTRODUCTION}
The galaxy merger rate over cosmic time is one of
the fundamental measures of the evolution of galaxies.
Galaxies and the dark matter halos they live in must
grow with time through mergers with other galaxies and
through the accretion of gas and dark matter from the
cosmic web. Over the past 10 billion years, the global
star-formation rate density has declined by a factor of
10 (e.g. Lilly et al. 1996; Madau et al. 1996; Hopkins
\& Beacom 2006) while the global stellar-mass density
has increased by a factor of two (e.g. Rudnick et al.
2003; Dickinson et al. 2003) . At the same time, massive
galaxies have been transformed from rapidly star-forming disk
galaxies into quiescent bulge-dominated galaxies hosting
super-massive black holes (e.g. Bell et al. 2004; Faber
et al. 2007; Brown et al. 2007). Galaxy mergers may
be an important process that drives galaxy assembly, rapid star-formation 
at early times, the accretion of gas onto central
super-massive black holes, and the formation of 
dispersion-dominated spheroids (e.g. Toomre 1977; White \&
Rees 1978;  Kauffmann, White, \& Guiderdoni 1993; 
Mihos \& Hernquist 1996; Sanders \& Mirabel 1996; Somerville et al.
2001, 2008; di Matteo et al. 2008; Hopkins et al. 2006, 2008). Because 
other physical processes are also at work, direct
observations of the galaxy mergers are needed to understand 
their global importance to galaxy evolution and
assembly.

In a cold-dark matter dominated universe, massive
structures are expected to grow hierarchically. Numerical 
simulations consistently predict that the dark matter
halo - halo merger rate per progenitor (or descendant)
halo at fixed halo mass changes rapidly with redshift
$\sim (1+z)^{2-3}$ (e.g. Gottl\"{o}ber, Klypin, \& Kratsvov 2001;
Fakhouri \& Ma 2008; Genel et al. 2009; Fakhouri, Ma, \& Boylan-Kolchin 2010). The dark matter 
merger rate scales with mass and mass ratio 
(Fakhouri \& Ma 2008; Fakhouri et al. 2010). More massive halos are
rarer, but have more frequent merger rates per halo. Minor 
mergers with mass ratios greater than 1:4 should be
much more common per halo than major mergers with
comparable mass halos. 

However, the theoretical predictions 
for the {\it galaxy} merger rate remain highly uncertain (e.g. Jogee et al.
2009; see Hopkins et al. 2010a for a review). The predicted
$(1 + z)^3$ evolution in the dark-matter halo merger rate
(per halo above a fixed total mass) does not automatically 
translate into a $(1 + z)^3$ evolution in the galaxy
merger rate (per galaxy above a fixed stellar mass) because 
there is not a simple connection between observed
galaxies and dark matter halos (e.g. Berrier et al. 2006; 
Hopkins et al. 2010a, Moster et al. 2010).   For example, the differences in the dark 
matter halo mass function and the galaxy stellar mass function at the high mass
and low mass ends naturally produces a discrepancy 
between the merger rates as a function of stellar and
dark-matter mass and mass ratio.  

Many galaxy evolution models predict that the evolution of major mergers of galaxies 
selected above a fixed stellar mass ($\sim 10^{10} M_{\odot}$) changes
more modestly with redshift ( $\sim (1+z)^{1-2}$), but these
can have an order of magnitude variation in their normalizations (e.g. 
Jogee et al. 2009;  Hopkins et al. 2010a). 
The largest source of theoretical uncertainty for semi-analytic
galaxy evolution models is the baryonic physics required to map galaxies onto
dark matter halos and sub-halos, and in turn, required to
match the observational selection of luminous merging
galaxies (Hopkins et al. 2010a and references therein).  
In contrast, semi-empirical models which map galaxies onto dark matter sub-halos
by matching the observed abundances of galaxies (per unit luminosity) to the 
abundances of sub-halos (per unit mass) give reasonably good predictions for clustering
statistics by adopting fairly small dispersions in the mapping function (e.g. Zheng et al. 2007, Conroy \& Wechsler 2009)
and give a only factor of two discrepancy in the predicted major merger rates (e.g. Hopkins et al. 2010b, Stewart et al. 2009a).

Despite more than a decade of work, measurements of the observed
galaxy merger rate and its evolution with redshift has
not converged. Current observations
of the fraction of bright/massive galaxies undergoing a
merger differ by an order of magnitude (Fig. 1), and
estimates of the evolution in this merger fraction of at
$0 < z < 1.5$ vary from weak or no evolution (e.g. Bundy
et al. 2004; Lin et al 2004; Lotz et al. 2008a; Jogee et al.
2009; Bundy et al. 2009; Shi et al. 2009; de Ravel et al.
2009; Robaina et al 2010) to strong evolution ( $\sim (1+z)^3$;
Le Fe\'{v}re et al. 2000; Conselice et al. 2009; Cassata et al.
2004; Kartaltepe et al. 2007; Rawat et al. 2008; Bridge
et al. 2010; L\'{o}pez-Sanjuan et al. 2009 ). It has been
difficult to understand the source of these discrepancies,
as the various studies often employ different criteria for
counting galaxy mergers and different selections for the
parent galaxy samples. 

These observational studies identify galaxy merger
candidates either as galaxies in close pairs (in projected
or real 3-D space) or galaxies with distorted morphologies 
(measured quantitatively or by visual inspection).
A key obstacle to the consistent measurement of the
galaxy merger rate has been the poorly constrained
timescales for detecting galaxy mergers selected by
different methods. Close pairs find galaxies before they
merge, while merger-induced morphological disturbances
can appear before, during, and after a galaxy merger.
Moreover, morphological studies use a variety of tracers 
to classify galaxies as mergers (such as $G-M_{20}$,
asymmetry, tidal tails), which have differing sensitivities to 
different merger properties such as mass ratio and
gas fraction. Thanks to new high-resolution hydrodynamical simulations
of galaxy mergers which model realistic light profiles including
the effects of star-formation and dust (e.g. Jonsson et al 2006), it
is now possible to estimate the timescales for detecting
galaxy mergers with a variety of close pair and morphological 
criteria (Lotz et al. 2008b; Lotz et al. 2010a, b).
These simulations also span a range of merger properties,
including progenitor mass, mass ratio, gas fraction, and
orbital parameters (Cox et al. 2006, 2008). Mass ratio and gas fraction are especially 
important for interpreting the merger fractions derived from morphological studies
(Lotz et al. 2010a, b).

The goal of this paper is to self-consistently determine
the observational galaxy merger rate at $z < 1.5$ from recent 
close pair and morphological studies. We weight the
merger timescales from individual high-resolution simulations 
by the expected distribution of merger properties (including
gas fraction and mass ratio) to determine 
the average observability timescale as a function of redshift for 
each method for finding galaxy mergers (close pairs, $G-M_{20}$, asymmetry). 
These new calculations of the average observability timescales are crucial 
for interpreting the observed galaxy merger fractions.   We re-analyze
the observations of galaxy mergers at $z < 1.5$ from the
recent literature and derive new estimates of the galaxy
merger rate. We also carefully consider the effects of
parent sample selection on the inferred evolution of the
galaxy merger rate. We conclude that the differences in the observed galaxy
merger fractions may be accounted for by the different observability
timescales,  different mass-ratio sensitivities, and different parent galaxy
selections.   When these differences in the methodology are properly 
accounted for by adopting our derived timescales, 
we are able to derive self-consistent estimates of the major 
and minor galaxy merger rate and its evolution. 

In \S2, we define the volume-averaged galaxy merger
rate, $\Gamma_{merg}$, as the co-moving number density of on-going
galaxy merger events per unit time, and the fractional
galaxy merger rate, $\Re_{merg}$, as the number of galaxy
merger events per unit time per selected galaxy. We
discuss how to calculate $\Gamma_{merg}$ and $\Re_{merg}$ 
using close pairs and disturbed morphologies. In \S3, we review the
recent literature estimates of the galaxy merger fraction 
using quantitative morphology ($G-M_{20}$, asymmetry) 
and close pairs of galaxies and discuss the importance 
of the parent sample selection. In \S4, we review
the results from the individual high-resolution galaxy
merger simulations. We present new calculations of the
cosmologically-averaged merger observability timescale
$\langle T_{obs}(z) \rangle$ for close pairs, $G-M_{20}$, 
and asymmetry, using the distribution of galaxy merger 
properties predicted by three different galaxy evolution 
models (Croton et al. 2006; Somerville et al. 2008; Stewart 
et al. 2009b). In \S5, we derive the fractional and volume-averaged 
major and minor galaxy merger rates for galaxy samples selected by 
stellar-mass, evolving luminosity, and constant co-moving number 
density at $z < 1.5$.  We compare these results to the predicted
major and minor galaxy merger rates. Throughout this
 work, we assume $\Omega_m = 0.3$, $\Omega_{\Lambda} = 0.7$, and $H_0 = 70$ 
 km s$^{-1}$ Mpc$^{-1}$, except for close pair projected separations 
 where $h = H_0 /100$ km s$^{-1}$ Mpc$^{-1}$. 

\section{CALCULATING THE GALAXY MERGER RATE}
The volume-averaged galaxy merger rate $\Gamma_{merg}$ is defined as the number
of on-going merger events per unit co-moving volume,
$\phi_{merg}$, divided by the time $T_{merg}$ for the merger to occur
from the initial encounter to the final coalescence:
\begin{equation}
  \Gamma_{merg} = \frac{\phi_{merg}}{ T_{merg}}
\end{equation} 
Note that $T_{merg}$, the time period between the time of the
initial gravitational encounter and coalescence (from ``not merging'' 
to ``merged'') is not well defined. 

More accurately, the number density of galaxies identified as
galaxy mergers $\phi_{merg}$ will depend on the average 
timescale $\langle T_{obs} \rangle$ during which the merger can be observed
given the method used to identify it, such that
\begin{equation}
\phi_{merg}' = \phi_{merg} \frac{\langle T_{obs} \rangle }{T_{merg}} 
\end{equation}
A galaxy merger may be identified at discontinuous stages 
(as in the case of close pairs),  therefore 
$T_{obs}$ is the sum of the observability windows. 

Thus, the galaxy merger rate $\Gamma$ can be calculated
from the observed number density of galaxy merger candidates
$\phi_{merg}'$ as follows:
\begin{equation}
\Gamma_{merg} = \frac{\phi_{merg}'}{T_{merg}} \frac{T_{merg}}{\langle T_{obs}\rangle} 
= \frac{\phi_{merg}'}{\langle T_{obs} \rangle}
\end{equation}

The majority of past merger calculations have assumed
$\langle T_{obs} \rangle$ to be a constant value ranging from $\sim$ 0.2 Gyr to 1
Gyr, without considering the differences in methodology,
the effect of merger parameters, or redshift-dependent
changes in the ability to detect mergers. In \S4, we present
detailed calculations of $\langle T_{obs} \rangle$ as a function of redshift for
different methods using theoretical models of galaxy interactions and 
the evolving distribution of galaxy merger properties.

Galaxy mergers can be directly identified as close
galaxy pairs that have a high probability of merging
within a short time. Galaxies with similar masses
that lie within a hundred kpc of each other and have
relative velocities less than a few hundred km s$^{-1}$ are
likely to merge within a few billion years. Individual close
galaxy pairs may be found by counting up the numbers of
galaxies with companions within a given projected separation 
$R_{proj}$ and within a relative velocity or redshift
range. However, spectroscopic determination of the relative 
velocities of close pairs is observationally expensive,
and prone to incompleteness due to slit/fiber collisions
and biased detection of emission-line galaxies (e.g. Lin
et al. 2004, 2008; de Ravel et al. 2009). Photometric
redshifts are often used as a proxy for spectroscopic velocity 
separations (e.g. Kartaltepe et al. 2007, Bundy
et al. 2009). However, even high precision photometric
redshifts ( $\frac{\delta z}{1+z} \sim 0.02$) are only accurate to 100-200 Mpc
along the line of sight. Therefore significant statistical corrections 
for false pairs are required for photometrically selected pair 
studies (e.g. Kartaltepe et al. 2007). Finally,
two-point angular correlation studies of large samples of
galaxies have also been used to determine the statistical
excess of galaxies within a given dark matter halo (e.g.
Masjedi et al. 2006, Robaina et al. 2010). These studies are 
less prone to line-of-sight projection issues, but cannot identify 
individual merging systems. 

Morphological disturbances also indicate 
a recent or on-going merger. Galaxy mergers experience strong
gravitational tides, triggered star-formation, and the redistribution 
of their stars and gas into a single relaxed galaxy. This 
gravitational rearrangement results in morphological distortions 
-- e.g. asymmetries, double nuclei, tidal tails -- which are 
detectable in high-resolution images. Such disturbances can be 
found through visual inspection by human classifiers, or by 
quantitative measurements of galaxy structures. Several different 
quantitative methods are commonly used to measure galaxy
morphology and classify galaxy mergers. Rotational
asymmetry ($A$) picks out large-scale high surface brightness 
asymmetric structures (e.g. Abraham et al. 1994;
Conselice, Bershady, \& Jangren 2000; Conselice 2003); the combination
of the Gini coefficient ($G$; Abraham et al. 2003; Lotz et al. 2004) and
second-order moment of the brightest 20\% of the light
($M_{20}$) selects galaxies with multiple bright nuclei (Lotz
et al. 2004). Both of these methods require high signal-to-noise 
and high spatial resolution images generally only
achievable with the Hubble Space Telescope ($HST$) at
$z > 0.3$, as well as a training set to distinguish `normal'
galaxies from merger candidates.

The merger fraction, $f_{merg}$, is the fraction of galaxies
identified as mergers for a given galaxy sample, and is
often presented instead of the number density of observed
merger events. Thus $\phi_{merg}'$ depends on both the merger
fraction and the co-moving number density of galaxies in
the selected sample, $n_{gal}$:
\begin{equation}
\phi_{merg}' = f_{merg} n_{gal} 
\end{equation}

For both morphologically-selected and close pair
merger candidates, a correction factor is applied to the
merger fraction to account for contamination from objects 
that are not mergers. For morphological merger
selection, this correction is applied prior to calculating
$f_{merg}$ based on visual inspection or a statistical correction. 
For close pairs, the fraction of pairs ($f_{pair}$) or the
fraction of galaxies in a pair ($N_c \sim 2 f_{pair}$) is multiplied
by this correction factor $C_{merg}$, such that
\begin{equation}
f_{merg} = C_{merg} f_{pair} \sim C_{merg} \frac{N_c}{2}
\end{equation}
We will adopt $C_{merg} = 0.6$ throughout this paper. 
Numerical simulations and empirical measurements suggest 
that $C_{merg}$ is $\sim$ 0.4$-$1.0 for close pairs selected with
projected separations $<$ 20$-$30 kpc $h^{-1}$ (Kitzbichler \&
White 2008, Patton \& Atfield 2008; see Bundy et al.
2009 for discussion).  It is worth noting that close pair samples
are likely to be biased towards objects in groups or
clusters (Barton et al. 2007), although simulations suggest that
the majority of these objects will merge (Kitzbichler \& White 2008). 

Multiple authors have estimated the fractional merger
rate $\Re_{merg}$ instead of $\Gamma_{merg}$ where
\begin{equation}
\Re_{merg} = \frac{f_{merg}}{ \langle T_{obs} \rangle} = \frac{C_{merg} f_{pair}}{\langle T_{obs} \rangle}
\end{equation}
(e.g. L\'{o}pez-Sanjuan et al. 2009, Bundy et al. 2009,
Bridge, Carlberg, \& Sullivan 2010, Conselice et al. 2009, Jogee et al.
2009). In principle, the fractional merger rate $\Re_{merg}$
circumvents variation in $n_{gal}$, which may change from field to
field with cosmic variance and is a factor of two or more
lower at $z \sim 1$ than $z \sim 0.2$ for fixed stellar mass or
passive luminosity evolution selection (Figure 2). Semi-analytic
cosmological simulations also have difficulty
simultaneously reproducing the correct galaxy number
densities at all mass scales, but may be more
robustly compared to observations for relative
trends such as $\Re_{merg}$.

$\Gamma_{merg}$ traces the number of merger events per co-moving 
volume above some mass/luminosity limit,
while $\Re_{merg}$ traces the number of merger events per
(bright/massive) galaxy. It is often implicitly assumed
that any evolution with redshift is not dependent on
$n_{gal}(z)$ or the galaxy selection either because the parent 
galaxy samples are selected in a uniform way or because 
the merger rate is not a strong function of galaxy
luminosity, mass, or number-density. However, several
studies have found that the fraction of mergers increases
significantly at fainter absolute magnitudes/lower stellar
masses (Bridge et al. 2010; de Ravel et al. 2009; Lin et
al. 2004; Bundy et al 2005), while other studies have found
increased merger fractions for more massive galaxies (Bundy et al. 2009; 
Conselice et al. 2003). If the observed merger fraction depends 
upon luminosity or stellar mass,  comparison of
different observational estimates of $\Gamma_{merg}$ or $\Re_{merg}$ 
will require careful consideration of the parent galaxy sample
selection criteria. We will return to this issue in \S3.5.

\begin{figure*}
\epsscale{1.2}
\plotone{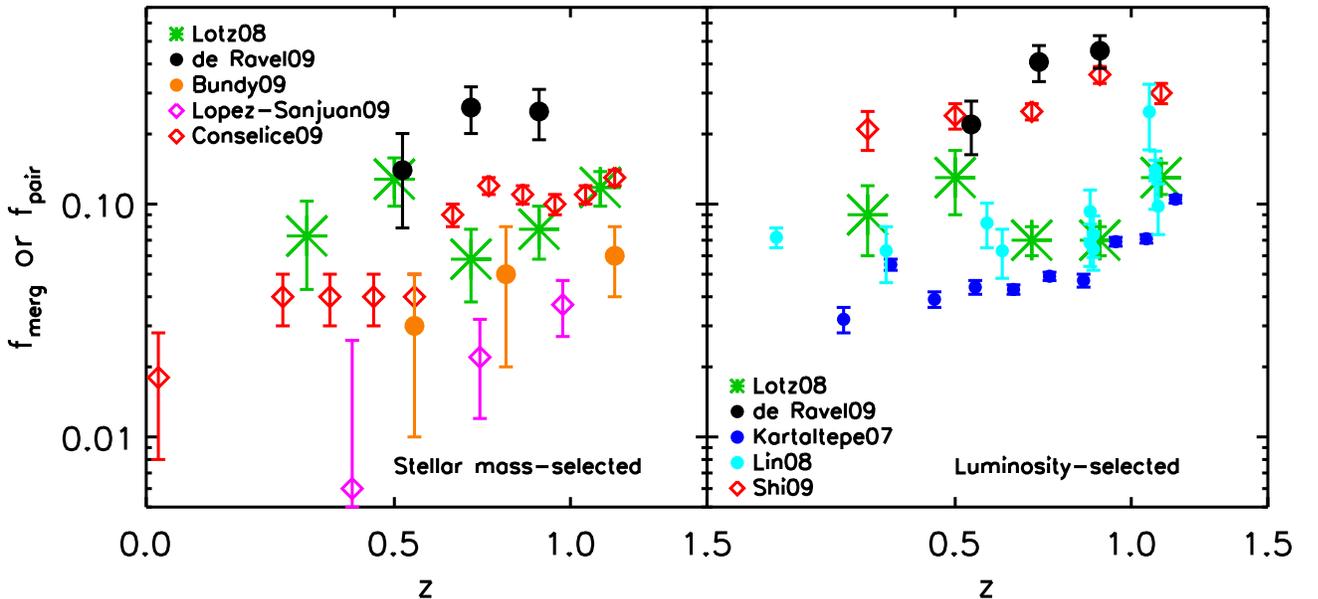}
\caption{Left: The galaxy merger fraction ($f_{merg}$) or close pair fraction ($f_{pair}$) v. redshift for samples selected by stellar mass
($M_{star} > 10^{10} M_{\odot}$). Right: Same, for samples selected with an evolving luminosity limit (see text). Large asterisks are $G-M_{20}$-selected
mergers, filled circles are for close pairs, and open diamonds are for asymmetric galaxies. For both stellar-mass and luminosity-selected
parent samples, $f_{merg}$ and $f_{pair}$ vary by a factor of 10 for different merger studies.}
\epsscale{1.}
\end{figure*}

\section{GALAXY MERGER OBSERVATIONS}
Despite the large number of galaxy merger studies,
there is little consensus on the galaxy merger rate or its
evolution with redshift at $z < 1.5$. In part, this is because
many studies have compared the galaxy merger fraction 
$f_{merg}(z)$ (Figure 1) or the galaxy merger
rate for galaxy merger samples selected with
different approaches and/or with different parent sample 
criteria. In this section, we review the selection of
galaxy mergers and their parent samples for a variety
of recent merger studies at $z < 1.5$. We will focus on
works that identify merger via quantitative morphology
($G - M_{20}$, $A$) or close pairs, but briefly discuss results
from other approaches. Almost all of these studies  
have calculated either the evolution in the merger fraction 
($f_{merg}$ or $f_{pair}$), thereby ignoring $\langle T_{obs}(z) \rangle$, 
or the merger rate $\Gamma_{merg}$ or $\Re_{merg}$ by 
assuming a constant $\langle T_{obs}\rangle$ with redshift.

\subsection{$G-M_{20}$-Selected Mergers}
One quantitative morphological approach to identifying 
galaxy mergers in high-resolution images has been
the $G - M_{20}$ method. The Gini coefficient $G$ is a 
measurement of the relative distribution of flux values in the
pixels associated with a galaxy, such that uniform surface 
brightness galaxies have low $G$ and galaxies with
very bright nuclei have high $G$ (Abraham et al. 2003;
Lotz et al. 2004). When $G$ is combined with $M_{20}$, the
second-order moment of the brightest 20\% of the light,
local spiral and elliptical galaxies follow a well-defined 
sequence and local mergers have higher $G$ and and higher
$M_{20}$ values (Lotz et al. 2004). Simulations of galaxy
mergers revealed that the $G - M_{20}$ technique primarily
identifies mergers during the first pass and final merger
(Lotz et al. 2008b) and is sensitive to mergers with baryonic 
mass ratios between 1:1 - 1:10 (Lotz et al. 2010b).
It has been noted by multiple authors that the merger
sample found by the $G - M_{20}$ technique is incomplete
(Kampczyk et al. 2007; Scarlata et al. 2007; Kartaltepe et al. 2010). 
This is consistent with the results from galaxy merger simulations, 
which predict that $G - M_{20}$ identifies merging
systems for only a short period while double nuclei are
evident (Lotz et al. 2008; Lotz et al. 2010a, b).

By applying the $G-M_{20}$ technique to high-resolution
$HST$ Advanced Camera for Surveys $V$ (F606W) and $I$
(F814W) images of the Extended Groth Strip (Davis et
al. 2007), Lotz et al. (2008a) found $f_{merg} = 10 \pm 2$\% 
for a sample of galaxies at $0.2 < z < 1.2$. This work concluded
that the galaxy merger fraction and rate evolved weakly
over this redshift range: $f_{merg}(z) \propto (1+z)^{0.23 \pm 1.03}$. 
The parent sample was selected to be brighter than $0.4L^*_B(z)$, 
assuming passive luminosity evolution of galaxies and
rest-frame B luminosity functions calculated for the Extended 
Groth Strip ($M_B < -18.94 - 1.3z$; Faber et al.
2007).    Only rest-frame $B$ morphologies were considered in
order to avoid biases associated with intrinsic morphological dependence
on rest-frame wavelength.  The merger fractions were corrected for visually-
and spectroscopically-identified false mergers  ($\sim$ 20$-$30\% of initial
merger candidates, similar to photometric redshift pair corrections).  
$G-M_{20}$ identified merger candidates with visually-identified artifacts or foreground
stars were removed from the sample.  Also, $G-M_{20}$ merger candidates with
multiple spectroscopic redshifts within a single DEEP2/DEIMOS slit were  
identified and used to statistically correct the resulting merger fractions (see
Lotz et al. 2008a for additional discussion.)    The 
statistical uncertainties associated with galaxies scattering 
from the main locus on ‘normal’ galaxies to merger-like 
$G - M_{20}$ values are also included in the merger fraction error. The $f_{merg}$ and $n_{gal}$ values from
Lotz et al. (2008a) are given in Table 2.

For this paper, we have also computed $f_{merg}(G-M_{20})$
for a parent sample selected above a fixed stellar mass
$M_{star} > 10^{10} M_{\odot}$ for the same $HST$ observations of the
Extended Groth Strip. (Stellar masses were calculated using Bruzual \& Charlot 2003 spectral
energy distributions and a Chabrier 2003 initial mass function; 
see Salim et al. 2009, 2007 for details).  We estimated $n_{gal}$
using the galaxy stellar mass functions v. redshift 
given by Ilbert et al. (2010) for the same redshift bins. We note that
these were derived using a different field (COSMOS), 
but are consistent with the galaxy stellar mass function 
computed for the Extended Groth Strip with larger redshift bins (Bundy et al.
2006).

\subsection{Asymmetric Galaxies}
Another commonly-used measure of morphological dis-
disturbance is the rotational asymmetry, $A$. Asymmetry measures
the strength of residuals when a galaxy’s profile is subtracted 
from its 180-degree rotated profile (Abraham et
al. 1994; Conselice et al. 2000). Merger simulations
suggest that high $A$ values may last for a longer period
of time than $G-M_{20}$ detected disturbances (Lotz et al.
2008b, 2010a,b). As we discuss in \S4, $A$ is highly 
sensitive to gas fraction and, for local gas fractions, probes
mergers with a different range of mass ratio than $G-M_{20}$.

Early studies of small deep $HST$ fields by Abraham et al. (1996) 
and Conselice et al. (2003, 2005) showed strong evolution 
in the fraction of asymmetric galaxies at $0 < z < 3$ with stellar masses
$> 10^{10} M_{\odot}$. Subsequent studies based on larger HST
surveys confirmed this initial result at $0 < z < 1.2$
(Cassata et al. 2005, Conselice et al. 2009, L\'{o}pez-
Sanjuan et al. 2009; but see Shi et al. 2009). For this
paper, we consider the measurements for three recent
studies: Conselice et al. (2009), L\'{o}pez-Sanjuan et al.
(2009) and Shi et al. (2009).   These studies compute $A$ 
in rest-frame $B$.  Low signal-to-noise images 
and redshift-dependent surface-brightness dimming
can produce strong biases in the asymmetry measurements. 
Each of these asymmetry studies makes different
assumptions about how to correct for these effects, and it is
therefore perhaps not surprising that they draw different conclusions about $f_{merg}(z)$.

Conselice et al. (2009) computed asymmetry fractions
for galaxies at $0.2 < z < 1.2$ selected from the COSMOS 
(Scoville et al. 2007 ), GEMS (Rix et al. 2004),
GOODS (Giavalisco et al. 2004) and AEGIS (Davis et
al. 2007) $HST$ surveys. They find that the fraction of
galaxies with stellar masses $> 10^{10} M_{\odot}$  that have high
asymmetry increases from $\sim$ 4\% at $z = 0.2$ to 14\% at
$z = 1.2$, with $f_{merg}(z) \propto (1+z)^{2.3 \pm 0.4}$. 
These asymmetry fractions are corrected for an assumed mean decrease in
asymmetry with redshift due to surface-brightness dimming 
($\langle \delta A \rangle = -0.05$ at $z = 1$). False merger candidates
scattered to high A values are visually identified and removed 
from the final asymmetric galaxy fractions, as are
galaxies with high $A$ but low clumpiness ($S$) values.

L\'{o}pez-Sanjuan et al. (2009) also compute the fraction of 
asymmetric galaxies in the GOODS fields selected
from a parent sample at $z < 1$ with stellar masses
$> 10^{10} M_{\odot}$. They find even stronger evolution in the
merger fraction $f_{merg}(z) \propto (1 + z)^{5.4 \pm 0.4}$. 
They apply a similar redshift-dependent
correction for surface-brightness dimming to the measured 
asymmetry values, but correct their asymmetry
values to the values expected to be observed at $z = 1$,
rather than $z = 0$ as Conselice et al. 2009 does. This
results in lower $f_{merg}$ values. Also, rather than visually 
inspecting the asymmetric merger candidates for false mergers, 
they employ a maximum-likelihood approach
to determine how many ‘normal’ galaxies are likely to
be scattered to high asymmetry values via large measurement 
errors and infer a high contamination rate ($>$ 50\%).
Consequently, their merger fractions are much lower than
Conselice et al. (2009), but follow similar evolutionary
trends with redshift.

Shi et al. (2009) also examine the fraction of asymmetric 
galaxies in $HST$ images of the GOODS fields. However, they 
select galaxies based on an evolving luminosity
cut ($M_B < -18.94-1.3z$) and apply different corrections for 
the effects of sky noise and surface-brightness dimming to the asymmetry measurements. Shi et al. (2009)
find that the fraction of asymmetric galaxies has not
evolved strongly at $z < 1$ with $f_{merg}(z) \propto (1+z)^{0.9 \pm 0.3}$, 
but find asymmetric fractions ∼ $2-3$ times higher than
Conselice et al. (2009). Shi et al. (2009) note that in
addition to a redshift-dependent surface-brightness dimming 
correction to the measured asymmetry values, correction 
based on the signal-to-noise of the image may be
required. Extremely deep images (i.e. the Hubble Ultra
Deep Field; Beckwith et al. 2006) can give asymmetry
values ∼ $0.05-0.10$ higher than lower signal-to-noise images 
of the same galaxy (Shi et al. 2009; Lotz et al.
2006). Based on extensive simulations of this effect, Shi
et al. (2009) correct the observed fractions of asymmetric
galaxies to that expected for high signal-to-noise observations 
in the local universe. No correction for falsely-identified merger 
candidates is applied, hence the Shi et al. (2009) results may be 
considered an upper limit to the fraction of asymmetric galaxies 
while the L\'{o}pez-Sanjuan et al. (2009) result may be considered a lower limit. It
is also possible that the different signal-to-noise thresholds for detecting 
asymmetry probe different processes (clumpy star-formation, minor mergers, 
major mergers).

\subsection{Close Pairs}
Numerous studies have attempted to measure the
galaxy merger rate by counting the numbers of galaxies 
with close companions. However, these studies often
adopt different criteria for the projected separation of
galaxies, the stellar mass (or luminosity) ratio of primary
and companion galaxy, and the stellar mass (or luminosity) 
range of the parent sample. We can account for
different projected separation criteria in our estimates of
the merger rate by modifying the dynamical timescale.
But without a priori knowledge of how the merger rate
depends on mass ratio and mass, we cannot compare
merger studies where close pairs are selected with different 
mass ratios or from samples with different limiting
masses. Here we describe the different criteria adopted
by different studies, and how these criteria are likely to
affect our derivation of the merger rate. We divide these
studies into those selected by stellar mass and those selected 
by rest-frame luminosities.    We calculate $\langle T_{obs}(z) \rangle$
for each of the close pair selection criteria in \S4, 
and assume $C_{merg} = 0.6$ for all samples when calculating the merger
rates in \S5. 

\subsubsection{Luminosity-selected pairs}
Lin et al. (2008) find weak evolution in the pair fraction 
between $0 < z < 1.1$, with $f_{pair} \propto (1 + z)^{0.4 \pm 0.2}$.
This work reanalyzes paired galaxies from the Team Keck
Treasury Redshift Survey (Wirth et al. 2004), DEEP2
Galaxy Redshift Survey (Davis et al. 2004; Lin et al.
2004), CNOC2 Survey (Yee et al. 2000; Patton et al.
2002), and Millennium Galaxy Survey (Liske et al. 2003;
de Propris et al. 2005) with a uniform selection criteria.
They require that both objects in the pair have spectroscopic 
redshifts, relative velocities $\delta V < 500$ km s$^{-1}$,
and projected separations $10 < R_{proj} < 30$ kpc h$^{-1}$. 
In order to restrict their sample to major mergers, they 
select only galaxies and their companion within a limited rest-frame 
$B$ luminosity range corresponding to a luminosity ratio 1:1 - 1:4. 
Based on the evolution of the rest-frame $B$ galaxy luminosity 
function with redshift observed by DEEP2
(Willmer et al. 2006; Faber et al. 2007), they assume
that the typical galaxy fades by 1.3$M_B$ per unit redshift,
and therefore have a redshift-dependent luminosity range
$-21 < M_B +1.3z < -19$. An additional complication is
the correction for paired galaxies where a spectroscopic
redshift is not obtained for one galaxy in the pair, either 
because of the spectroscopic survey sampling (not
observed) or incompleteness (not measurable). This correction 
is a function of color, magnitude, and redshift,
and therefore is not trivial to compute. We give $N_c$ (uncorrected) 
values and completeness-corrected 
pair fractions $f_{pair}$ from Lin et al. (2008) in Table 1.

de Ravel et al. (2009) recently completed a similar
study, using galaxies with spectroscopic redshifts $0.5 <
z < 0.9$ from the VIRMOS VLT Deep Survey (VVDS).
They find that the evolution in the pair fraction depends
on the luminosity of the parent sample, with brighter
pairs showing less evolution ($f_{pair}  \propto (1+z)^{1.5 \pm 0.7}$) than
fainter pairs ($f_{pair} \propto (1 + z)^{4.73 \pm 2.01}$). 
(See also L\'{o}pez-Sanjuan et al. 2011 for an analysis of minor
mergers with the same data.)
We will examine only the bright sample results, as this selection
($M_B < -18.77 - 1.1z$) is similar to the Lin et al. (2008)
study. They adopt a similar luminosity ratio for the
paired galaxies (1:1 - 1:4) to the Lin et al. 2008 study.
They also compute the pair fraction $f_{pair}$ at several different projected separations. 
We use their values for $R_{proj} < 100$ kpc $h^{-1}$ and 
$\delta V \leq 500$ km s$^{-1}$ because these have the 
smallest statistical errors and give merger rates consistent 
with smaller projected separation. The number density of
 galaxies was computed using the VVDS rest-frame $B$ 
luminosity functions (Ilbert et al. 2005).

Kartaltepe et al. (2007) used photometric redshifts to
select objects at $0.2 < z < 1.2$ with close companions
from deep $K$-band imaging of the COSMOS field. They
find that the pair fraction evolves strongly with redshift
$f_{pair} \propto (1 + z)^{3.1 \pm 0.1}$, in rough agreement with a later
study by Rawat et al. (2008) of $J$-band selected pairs
in the Chandra Deep Field South. To minimize the contamination 
from false pairs, they applied a stricter projected separation 
criteria of $5 < R_{proj} < 20$ kpc $h^{-1}$ and
$\delta z_{phot} \leq 0.05$. They also selected pairs of galaxies where
both objects were brighter than a fixed absolute magnitude 
$M_V = -19.8$ which corresponds to $L^*_V$ at $z = 0$.  They
correct the pair fractions for contamination rates for each redshift bin 
($\sim 20-30$\%).   The range of mass ratios is not well-defined, but $> L^*$
galaxies are rare, thus the majority of pairs will have
luminosity ratios between 1:1 and 1:4.

Because their luminosity selection does not evolve with
redshift, the Kartaltepe study probes a fainter parent
population of galaxies at $z \sim 1$ and brighter parent 
population at $z \sim 0.2$ relative to the de Ravel and Lin et
al. spectroscopic pair studies and the Lotz et al. (2008a)
$G-M_{20}$ study. This makes it difficult to directly compare 
the Kartaltepe et al. (2007) results to other studies 
which adopt evolving luminosity selection. They do consider 
the evolution of $f_{pair}$ when an evolving luminosity 
cut $M_V < -19.8 + 1.0z$ is adopted, and find that
their conclusions about the evolution of $f_{pair}(z)$ do not
change. In this work, we will use the COSMOS pair fractions 
selected with $M_V < -19.8 - 1.0z$ (J. Kartaltepe,
private communication) to compare with the other merger studies.
Because no published rest-frame $V$ luminosity function
is available for the COSMOS field, we estimate $n_{gal}$ for
$M_V < -19.8 - 1.0z$ from the Ilbert et al. (2005) rest-frame
$V$-band luminosity function of the VVDS field. We will
return to the issue of parent sample selection and its effects 
on the evolution in the observed galaxy merger rate
in \S3.5 .

Patton \& Atfield (2008) select $z \sim 0.05$ close pairs
with spectroscopic redshifts from the Sloan Digital Sky
Survey. These are required to have $\delta V < 500$ km s$^{-1}$,
$5 < R_{proj} < 20$ kpc $h^{-1}$, and rest-frame $r$-band 
luminosity ratios between 1:1 and 1:2. The parent sample is
drawn from galaxies with $-22 < M_r < -18$. They derive
a completeness-corrected value of $N_c = 0.021 \pm 0.001$.
Patton \& Atfield argue that between 50-80\% of these
pairs are line-of-sight projections, higher than the 40\%
($C_{merg} \sim 0.6$) assumed here and in other works.

\subsubsection{Stellar-mass selected pairs}
In order to avoid assumptions about galaxy rest-frame
luminosity evolution, Bundy et al. (2009) select galaxy
pairs in the GOODS fields based on a fixed stellar mass
and find that the pair fraction does not evolve strongly
at $z < 1.2$, with $f_{pair} \propto (1 + z)^{1.6 \pm 1.6}$. They compute
stellar masses by fitting the galaxies' spectral energy distributions 
($0.4-2$ $\mu$m) with Bruzual \& Charlot (2003)
stellar population models, assuming a Chabrier (2003)
stellar initial mass function. The galaxy sample is selected 
above a fixed stellar mass limit of $M_{star} \geq 10^{10} M_{\odot}$. 
Close pairs are selected to be within $5 < R_{proj} < 20$ kpc
h$^{-1}$ and with stellar mass ratios between 1:1 and 1:4.
Bundy et al. (2009) includes paired galaxies with both
spectroscopic and photometric redshifts where 
$\delta z^2 < \sigma^2_{z,primary} + \sigma^2_{z, satellite}$
and the typical photometric redshift error 
$\sigma_z < 0.08 (1+z)$. We estimate the number density
of GOODS galaxies with stellar masses $> 10^{10} M_{\odot}$ 
from Bundy et al. (2005).

In addition to their luminosity-based selection, de
Ravel et al. (2009) also select spectroscopic VVDS
galaxy pairs by stellar mass. The derived pair fraction
evolution depends on the stellar mass limit of the sample,
with more massive pairs showing weak evolution 
($f_{pair} \propto (1+z)^{2.04 \pm 1.65}$),  
consistent with the Bundy et al. (2009) results. 
Their stellar masses are computed with similar
population synthesis models and multi-wavelength photometric 
data to the Bundy et al. (2009). de Ravel et
al. (2009) assumes a modest evolution of stellar mass
 and selects a parent sample with
an evolving stellar mass limit log[$M_{star}$/$10^{10} M_{\odot}$] $> 0.187 z $.  
As for the luminosity selected sample, we
use de Ravel et al.~'s spectroscopic pairs with
$R_{proj} < 100$ kpc $h^{-1}$ and $\delta V \leq$500 km s$^{-1}$. 
Galaxy pairs are required to have stellar mass ratio between 1:1
and 1:4. We estimate the number density of galaxies
from the Pozzetti et al. (2007) $K$-selected stellar mass
functions computed for the VVDS.

\subsection{Visually Classified Mergers}
The visual classification of morphologically disturbed
galaxies has a long history (e.g. Hubble 1926), and has increased 
in scale and sophistication.  Several recent studies
have estimated the galaxy merger fraction and rate using
large samples of visually-classified mergers at $z < 1.5$, 
including the Galaxy Zoo project (Darg et al. 2010; also
Jogee et al. 2009; Bridge et al. 2010; Kartaltepe et al. 2010). However, studies
of the evolution of visually-classified mergers also reach
contradictory conclusions.

\begin{figure}
\epsscale{1.2}
\plotone{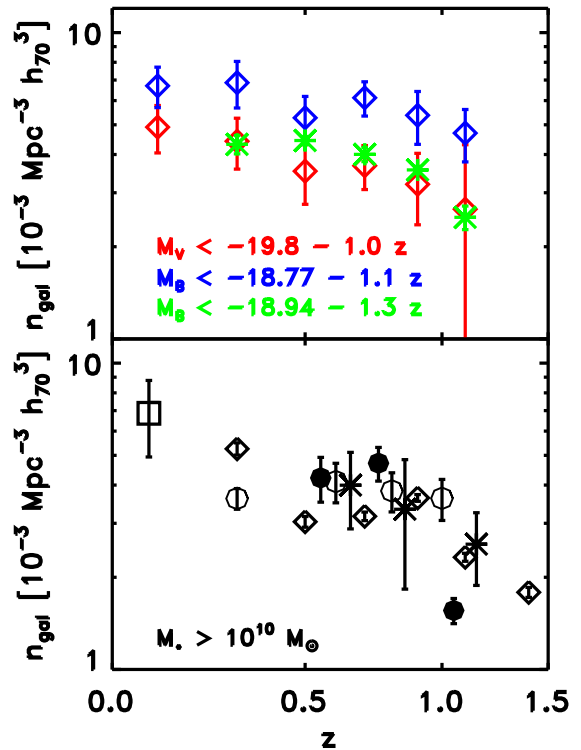}
\caption{ Top: Number of galaxies per co-moving unit volume
$n_{gal}$ v. redshift for the evolving luminosity selection calculated
from Ilbert et al. 2005 (VVDS $V$-band selected, red diamonds;
VVDS $B$-band selected blue diamonds) and Faber et al. 2007
(DEEP2 $B$-band selected, green asterisks); Bottom: $n_{gal}(z)$ for
galaxies selected with $M_{star} > 10^{10} M_{\odot}$ from Bell et al. 2003
(2MASS, black square), Bundy et al. 2005 (GOODS, black filled
circles), Ilbert et al. 2009 (black diamonds), Pozzetti et al. 2007
(VVDS, open circles), and Bundy et al 2006 (DEEP2, black asterisks).}
\epsscale{1.}
\end{figure}

Jogee et al. (2009) visually classified galaxies with stellar 
masses $> 2.5 \times 10^{10} M_{\odot}$ at $0.24 < z < 0.80$ 
in $V$-band $HST ACS$ images from the GEMS survey. 
They identify mergers as objects with asymmetries, shells, double
nuclei, or tidal tails but exclude obvious close or interacting pairs. 
They find that $\sim$ 9\% of the sample were
visually-disturbed with little evolution between $z \sim 0.2$
and $z \sim 0.8$. Between 1-4\% of the sample are classified
as ‘major’ mergers and ∼ 4-8\% are classified as ‘minor’
mergers. They conclude that while most massive galaxies
have undergone a major or minor merger over the past 7
Gyr, visually-disturbed galaxies account for only 30\% of
the global star-formation at those epochs.

Bridge et al.  (2010) also visually identify galaxy 
mergers at $z < 1$, but reach fairly different
conclusions. The Bridge et al. (2010) study is based
upon very deep ground-based images from the CFHT
Legacy Survey, therefore has worse spatial resolution
than $HST$-based studies but reaches comparable limiting 
surface brightnesses over a much wider area. They
identify merging and interacting galaxies via extended
tidal tails and bridges for a sample of $\sim$ 27,000 galaxies
with $i < 22$ Vega mag. They find the merger fraction of
galaxies with $i < 22$ and stellar mass $> 3 \times 10^9 M_{\odot}$ 
increases rapidly from $\sim$ 4\% at $z \sim 0.4$ to $\sim$ 19\% 
at $z \sim 1$, with an evolution $\propto (1+z)^{2.25 \pm 0.24}$.

We will not attempt to incorporate these results into
our study here. Visual classification of galaxy mergers
is  qualitative by nature, and the exact merger identification 
criteria applied for each study depends on the human
classifiers as well as the spatial resolution and depth of
the imaging data. Therefore visual classification ‘observability’ timescales
are unique to each study and less straightforward to
calculate than for the quantitative and easily reproducible (but
possibly less sensitive; Kartaltepe et al. 2010) methods that we use in this work.

\subsection{The Importance of Sample Selection}
Often the merger fraction $f_{merg}$ is measured over a
range of redshifts where the galaxy mass and luminosity
functions change significantly (e.g. Faber et al. 2007; 
Ilbert et al. 2010). If the galaxy merger rate is a function
of stellar mass or luminosity (e.g. Lin et al. 2004; Bundy
et al. 2005; de Ravel et al. 2009; Bridge et al. 2010),
then one must be careful about comparing studies with
different parent selection criteria. In an attempt to sample 
similar galaxies over a wide redshift range, the parent
galaxy sample may be selected to be more massive than
a fixed stellar mass (e.g. Bundy et al. 2009, Conselice et
al. 2009, de Ravel et al. 2009, Jogee et al. 2009, Bridge
et al. 2010) or by adopting an evolving luminosity selection 
based on a ‘passive luminosity evolution’ (PLE)
model in which galaxies of a fixed stellar mass are fainter
at lower redshift due to passive aging of their stellar 
populations (e.g. Lotz et al. 2008, Lin et al. 2008, de Ravel
et al. 2009, Shi et al. 2009).

We compare the number densities of galaxies selected
with passive luminosity evolution assumptions (Fig. 2,
upper panel) and with a fixed stellar
mass (Fig. 2, lower panel). The number densities
for the PLE cuts $M_B < -18.94 - 1.3z$
and $M_V < -19.8 - 1.0z$ and the fixed stellar mass cut
$M_{star} \geq 10^{10} M_{\odot}$ are similar at $z < 1$, suggesting that the
evolving luminosity selections of Lin et al. 2008, Lotz et
al. 2008, Kartaltepe et al 2007, Shi et al. 2009, and the
fixed stellar-mass selection of Bundy et al. 2009, Conselice
et al. 2009, de Ravel et al. 2009, and L\'{o}pez-Sanjuan
et al. 2009 probe similar galaxy populations. However,
the number density of galaxies selected by the de Ravel et al.
(2009) $M_B < -18.77-1.1z$ cut is higher at higher $z$ than
the other selections. When a fixed luminosity (`no evolution')
 cut is adopted, as in Kartaltepe et al. 2007 and
Rawat et al. 2008, the differences in the parent sample at
$z \sim 1$ are even more significant, which could explain the
difference in the derived evolution of the mergers. We
will return to this issue in \S5.4.

An additional complication is that merging galaxies increase in 
stellar mass and luminosity as they merge, as well as undergo shorter-lived
starbursts.  Therefore late-stage mergers and merger remnants will be more
massive and luminous than their progenitors.    Morphological disturbances
are often late stage mergers whose nuclei are counted as a single objects, 
while close pairs are early-stage mergers (e.g. Lotz et al. 2008).  Thus  morphologically-disturbed
mergers at a given mass/luminosity are drawn from a less-massive progenitor
population than close pairs selected at the same mass/luminosity limit.   
In the worst case of equal-mass mergers,  the morphologically-disturbed mergers
will have progenitors 50\% less massive and up to a magnitude fainter than 
close pairs.   The inferred merger rate from morphologically-disturbed objects could 
be biased to higher values than the close-pair merger rate if the number density of merger
progenitors increases with lower stellar mass.     However,  as we discuss in the
next section,  both $G-M_{20}$ and $A$ are sensitive to minor mergers as well as major mergers. 
Therefore the difference between the typical (primary) progenitor's and the remnant's stellar mass
is less than 25\% for morphologically-selected samples, assuming a typical merger ratio 1:4.  
A related issue is short-lived luminosity brightening
of both close pairs and morphologically-disturbed galaxies from merger-induced starburst. 
This effect could bias luminosity-selected samples similarly,  and give different merger rates
than samples selected by stellar mass.   However, merger simulations suggest that luminosity brightening
in the rest-frame optical is mitigated by dust obscuration of the starbursts (Jonsson et al. 2006 ). 
 In \S5, we find little difference between the merger rates for luminosity and stellar-mass selected samples,
implying that luminosity brightening does not introduce signification biases to the merger rate. 

Finally, it is also important to note that the standard selection of galaxy
 sample for galaxy merger studies (e.g.
above a fixed stellar mass) is not well matched to the
way that dark matter halos are selected from simulations 
for dark matter halo merger studies (e.g. above a
fixed halo mass). We know that typical blue galaxies are not
evolving purely passively, but form a significant number
of new stars at $z < 1$ (e.g. Lilly et al. 1996 ; Noeske
et al. 2007). Therefore the relationship between stellar
mass and host dark matter halo mass also evolves with
redshift (Zheng, Coil, \& Zehavi 2007; Conroy \& Wechsler
2009; Moster et al. 2010). Given that the co-moving number density of galaxies with
stellar masses $> 10^{10} M_{\odot}$ or selected based 
on PLE assumptions is as much as a factor of four lower at $z \sim 1.5$ 
than $z \sim 0$, it is clear that neither of these galaxy samples select the
progenitor and descendant galaxies across the range of
redshifts.

Galaxy samples selected with a constant number density 
may do a better job of matching descendant-progenitor 
galaxies over a range of redshifts (van Dokkum et al. 2010; 
Papovich et al. 2010), and matching galaxies to constant mass 
halos (at $M_{halo} \sim 10^{11-12} M_{\odot}$; Zheng et al. 2007).
This is not a perfect selection criterion, as it assumes
galaxy mergers do not destroy significant numbers of galaxies
and that stochasticity in the luminosity/mass evolution
of galaxies does not strongly affect the sample selection.
In \S5.4, we present galaxy merger rates for parent galaxy
samples selected with a roughly constant number density
at $z < 1.5$, and compare these to the rates derived for
the fixed stellar mass and PLE samples.

\section{MERGER OBSERVABILITY TIMESCALES}
Knowledge of the average merger observability
timescale $\langle T_{obs} \rangle$ is crucial for calculating the galaxy
merger rate. This timescale will depend upon the method
used to select galaxy mergers, as close pairs pick out different 
merger stages from objects with disturbed morphologies. 
For a given merger system, the individual observability 
timescale will also depend on the parameters
of the merger, such as the initial galaxy masses, morphologies and 
gas fractions, the merger mass ratio, the orbit,
and the relative orientation of the merging galaxies. The
initial properties of the merging system are difficult to reliably 
recover from observations of an individual merging
system, particularly after the merger has progressed to
late stages. Moreover, the observed merger population is
a mixture of merger events with a broad distribution of
parameters which may be evolving with redshift. Therefore 
the mean observability timescale $\langle T_{obs} \rangle$ 
given in Eqn. 3 should be treated as a cosmologically-averaged 
observability timescale weighted by the distribution of merger
parameters at each epoch.

Because the observations of a given galaxy merger
are essentially an instantaneous snapshot at a particular
merger stage, we cannot know how long that merger
event will exhibit disturbed morphology or other obvious
merger indicators. And, depending on the merger stage
and viewing angle, it can be difficult to determine its initial conditions. 
Therefore, the best way to determine how
the morphology and projected separation of a merger
progresses and how this depends on the merger conditions 
is by studying a large number of high-resolution numerical 
simulations of individual mergers where the initial
merger conditions are systematically explored. Also, we
do not yet have good observational constraints on the
distribution and evolution of galaxy merger parameters
such as gas fraction or mass ratio. Thus we will use theoretical 
predictions from three different global galaxy evolution models
of the distribution of galaxy properties and their evolution to 
constrain the distribution of merger parameters
with redshift.

In this section, we summarize the results of recent calculations
 of the individual observability timescales from
a large suite of high-resolution disk-disk galaxy merger
simulations for the $G-M_{20}$, $A$, and close pair methods.
Then we examine the predicted distributions of two key
merger parameters -- baryonic gas fraction and mass ratio
-- and their evolution with redshift from three different
cosmological-scale galaxy evolution models (Somerville
et al. 2008; Croton et al. 2006; Stewart et al 2009b).
Finally, we use these distributions to weight the individual 
observability timescales and derive the average 
observability timescales as a function of redshift $\langle T_{obs}(z) \rangle$
for each galaxy evolution model and approach to detecting 
galaxy mergers. We compare the predictions of the
different galaxy evolution models and discuss the uncertainties 
associated with $\langle T_{obs}(z) \rangle$. The derived $\langle T_{obs}(z) \rangle$
are applied to the observed merger fractions described in
\S3 to compute merger rates in \S5.

\begin{figure*}
\plotone{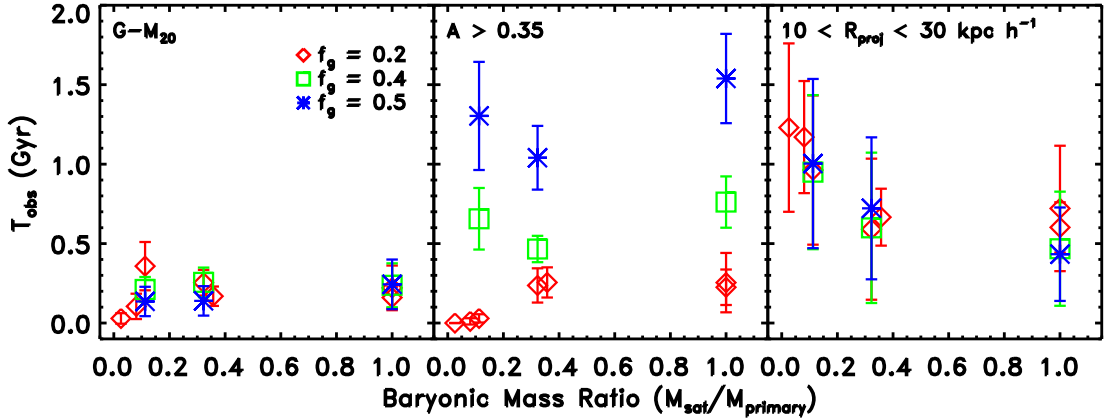}
\caption{The observability timescales $T_{obs}$ for individual prograde-prograde disk-disk galaxy merger simulations v. baryonic mass ratio
($M_{primary}/M_{satellite}$) from Lotz et al. 2010a,b. The simulations were run with three different baryonic gas fractions $f_{gas} = 0.2$ (red
diamonds: G3G3Pt, G3G2Pt, G3G1Pt, G3G0Pt, G2G2Pt, G2G1Pt, G2G0Pt), 0.4 (green squares: G3gf1G3gf1, G3gf1G2, G3gf1G1) , and
0.5 (blue asterisks: G3gf2G3gf2, G3gf2G2, G3gf2G1).}
\end{figure*} 

\begin{figure*}
\plotone{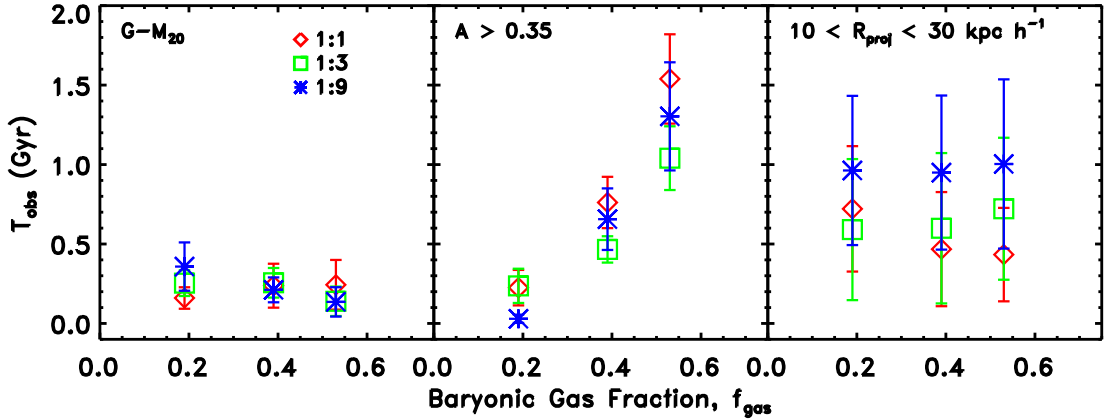}
\caption{ The observability timescales $T_{obs}$ for individual prograde-prograde disk-disk galaxy merger simulations v. baryonic gas fraction
$f_{gas}$ from Lotz et al. 2010a,b. Here, the points designate three different baryonic mass ratios $M_{satellite}/M_{primary}$ ∼ 1:1(red diamonds:
G3G3Pt, G3gf1G3gf1, G3gf2G3gf2), 3:1 (green squares: G3G2Pt, G3gf1G2, G3gf2G2), and 9:1 (blue asterisks: G3G1, G3gf1G1,
G3gf2G1).}
\end{figure*}

\subsection{Merger Timescales from Dusty Interacting Galaxy GADGET/SUNRISE Simulations}
The individual merger observability timescales $T_{obs}$
were calculated for a large suite of high-resolution N-body/hydrodynamical 
galaxy merger simulations in Lotz et al. 2008b, 2010a, b (also known as DIGGSS,
Dusty Interacting Galaxy {\sc GADGET/SUNRISE} Simulations; 
see http://archive.stsci.edu/prepds/diggss for simulation images and morphology measurements). 
Each {\sc GADGET} simulation tracks the merger of two disk
galaxies within a box of $\sim$ (200 kpc)$^3$ with a spatial resolution
 $\sim$ 100 pc and a particle mass  $\sim 10^5 M_{\odot}$, over a several billion
year period in $\sim$ 50 Myr timesteps. Cold gas
is converted into stars assuming the Kennicutt-Schmidt
relation (Kennicutt 1998). The effects of feedback from
supernovae and stellar winds using a sub-resolution effective equation of 
state model is included, but feedback
from active galactic nuclei (e.g. Di Matteo et al. 2008)
is not. The full suite of simulations span a range of initial 
galaxy masses and mass ratios, gas fractions, and
orientations and orbital parameters and are described in
detail in Cox et al. 2006, 2008.   The DIGGSS simulations 
do not include any $f_{gas} < 0.2$ simulations or
spheroidal merger simulations. 

The  {\sc GADGET} predictions for the ages, metallicities, and
3-D distribution of stars and gas for each simulation and
time-step were run through the Monte-Carlo radiative
transfer code {\sc SUNRISE} to simulate realistic 
ultraviolet-optical-infrared images, including the effects of star-formation, 
dust, and viewing angle. The details of the
{\sc SUNRISE} code and the predicted effects of dust on the
integrated spectral energy distributions are described in
Jonsson 2006,  Jonsson, Groves, \& Cox 2010. In Lotz et al. (2008b, 2010a, 2010b),
the simulated broad-band SDSS $g$ images 
were used to calculate the time during which each
individual galaxy merger simulation would be counted
as a ‘merger candidate’ by the quantitative morphology
$G - M_{20}$ and asymmetry methods.   We also calculate
close pair timescales for several different projected separation
criteria ($5 < R_{proj} < 20$, $10 < R_{proj} < 30$, $10 < R_{proj} < 50$, and
$10 < R_{proj} < 100$ kpc $h^{-1}$);  all simulations have merging
galaxies with relative velocities less than 500 km s$^{-1}$.  
Note that for the close pair and $G-M_{20}$ methods, the observability
windows are not contiguous in time and therefore the observability
time is the sum of the times when the merger is selected, e.g. as
a close pair before and after the first pass. 

We found that, for a given method, $T_{obs}$ depended most
on the mass ratio of the merger (Lotz et al. 2010a) and on
the gas fractions of initial galaxies (Lotz et al. 2010b).
Orbital parameters (eccentricity, impact parameters) and total
mass of the initial galaxies had little effect on $T_{obs}$ for morphological
disturbances. 
We will refer to the baryonic mass ratio of the merger
$M_{satellite}/M_{primary}$, where major mergers have baryonic
mass ratios between 1:1 and 1:4 and minor mergers have
baryonic mass ratios less than 1:4. We have chosen to
characterize the merger mass ratio by the initial baryonic 
mass ratio rather than total or stellar mass ratio
as a compromise between observational and theoretical
limitations. The total mass ratio of mergers is difficult
to estimate (or even define) observationally because this
requires a dynamical estimate of mass associated with
each interacting galaxy. On the other hand, the stellar
masses and stellar mass ratios predicted by cosmological
galaxy evolution models can be quite sensitive to the uncertain
 physics of star formation and feedback (see, for
example, Benson et al. 2003).  As a result, a galaxy merger with a total
mass ratio of 1:3 may correspond to a stellar mass ratio of 
anywhere from 1:10 to 1:2 (Stewart 2009a).  While baryonic mass ratios 
suffer the same complication, one may expect the ratio of 
baryonic-to-total mass to vary less strongly with redshift, since higher gas fractions at earlier 
times help offset the proportionately smaller stellar-to-total masses of $\sim L^*$ galaxies.

Likewise, we characterize the merger gas fraction as
the initial fraction of the baryons in cold gas. We define the
initial baryonic gas fraction $f_{gas}$ of the merger as
\begin{equation}
f_{gas} = \frac{(M_{gas,1} +M_{gas,2})}{(M_{gas,1} +M_{gas,2} +M_{star,1} +M_{star,2})} 
\end{equation}
This is the average gas fraction of the system prior to
the merger, when the galaxies first encounter each other.
Because these simulations do not accrete additional gas
and gas is converted into stars, the baryonic gas fraction
at the final merger is lower than the initial value.
For this work, we use a sub-set of DIGGSS which
have parabolic orbits, tilted prograde-prograde orientations 
and initial galaxy parameters tuned to match SDSS galaxies, 
but span a range of galaxy mass ($M_{star} \sim  1 \times 10^9 - 
5 \times 10^{10} M_{\odot}$ ) and baryonic mass ratios (1:1
- 1:39): G3G3Pt, G3G2Pt, G3G1Pt, G3G0Pt, G2G2Pt,
G2G1Pt, G2G0Pt. These simulations have gas fractions
tuned to local galaxies, with the initial baryonic gas fractions 
$f_{gas} \sim 0.2$. We also use the subset of DIGGSS
which have the same parabolic orbits, tilted prograde-
prograde orientations, and mass ratios but higher gas
fractions ($f_{gas} \sim$ 0.4, 0.5) : G3gf1G3gf1, G3gf1G2,
G3gf1G1, G3gf2G3gf2, G3gf2G2, G3gf2G1. See Lotz et
al. (2010a, b) and Cox et al. (2008) for more details of these
simulations.

In Figures 3 and 4, we show how the individual observability 
timescales $T_{obs}$ for $G-M_{20}$, $A$, and close pairs
with $10 < R_{proj} < 30$ kpc $h^{-1}$ depend on baryonic mass
ratio and $f_{gas}$.  Note that these are averaged over 11 different
viewing angles (see Lotz et al. 2008b).  The error-bars on $T_{obs}$ in Figures 3 and
4 are the standard deviation with viewing angle of $T_{obs}$
for each simulation. $G-M_{20}$ and close pair observability 
timescales are largely independent of gas fraction but
are sensitive to mass ratio, while asymmetry $A$ is sensitive to
both gas fraction and mass ratio. $G-M_{20}$ detects mergers 
with baryonic mass ratios between 1:1 and at least
1:10, and does not show a strong correlation of $T_{obs}$ with
baryonic mass ratio above 1:10. Mergers are not found
with the $G-M_{20}$ criteria for the two simulations with
baryonic mass ratios less than 1:10 (G2G0Pt, G3G0Pt), therefore
we conclude that $T_{obs} \sim 0$ for mergers at these mass ratios.
The simulations imply that $G-M_{20}$ detects mergers at the
stage when two bright nuclei are enclosed in a common envelope, 
and that satellites with masses greater than a tenth of the primary
galaxies are bright enough to be detected  (see Lotz et al. 2008a, Lotz 
et al 2010a for discussion). 

Dynamical friction and the effects of projection largely drive
the close pair timescales,  hence their insensitivity to $f_{gas}$. 
The close pair timescales are correlated with mass and  mass ratio such 
that minor mergers appear as close pairs for a
longer period of time than major mergers, as do lower-mass major mergers
(Table 5 in Lotz et al. 2010a).     

The behavior of $A$ observability timescales is more
complicated because of their dependence on both mass ratio 
and gas fraction.  The simulations suggest that high
asymmetries are the result of large-scale disturbances, 
including bright star-forming tidal tails and dust lanes,  thus
asymmetry is higher for more gas-rich mergers which are
more likely to form strong tidal features (see Lotz et al. 2010b). 
At gas fractions expected for typical 
local disk galaxies ($f_{gas} \sim 0.2$), $A$ detects primarily
major mergers with baryonic mass ratios between 1:1
and 1:4, and has $T_{obs} \sim 0$ for more minor mergers with
baryonic mass ratios 1:10 (Figure 3, red points, center
panel). But for simulations with $f_{gas} \ge 0.4$, $A$ detects
both 1:1-1:4 major mergers and 1:10 minor mergers 
(Figure 3, center panel, green and blue points). The 
observability timescale for $A$ is strongly correlated with $f_{gas}$
(Figure 4, center panel), with high gas fraction mergers
showing high asymmetries for significantly longer periods
of time than low gas fraction mergers. Therefore knowledge 
of both the merger gas fraction and mass ratio distributions is
required to estimate $\langle T_{obs} \rangle$ for asymmetry.

For this work, we will ignore the effect of the orbital parameters 
and relative orientations of the merging galaxies on the observability
 timescales.  All of the simulation timescales adopted here are for the
G-series prograde-prograde orientations, with $e =0.95$
and  pericentric distances $\sim 0.01-0.05$ times the viral radii of the
progenitors. In Lotz et al. (2008b) and (2010a), we found that different orbits and 
orientations had a weak effect on the observability timescales
for $G - M_{20}$ and $A$. However, galaxies merging on circular 
orbits or with large impact parameters were identified as 
close pairs for ∼ 15-40\% longer than parabolic
orbits with smaller impact parameters, depending on the
projected separation requirements. Large-scale numerical 
simulations find that the majority of dark-matter
halo mergers are parabolic as opposed to circular 
(Khochfar \& Burkert 2006; Benson 2005; Wetzel 2010), with only modest
evolution in the typical orbital eccentricity out to $z \sim 1.5$ (Wetzel 2010).
Our adopted $e$ is consistent with the mean value for dark matter halo -satellite
mergers from Wetzel (2010;  $e \sim 0.85$),  while our pericentric distances 
are smaller than mean ($\sim$ 0.13 $R_{vir}$).   If true pericentric distances are
significantly larger than our assumptions here,  our simulations imply that this  
could systematically increase the merger timescales (and decrease merger rates) by 
up to 10-40\% for close pairs and a factor of 3-4 for $G-M_{20}$ selected 
major mergers (Lotz et al. 2008b).  

\begin{figure*}
\plotone{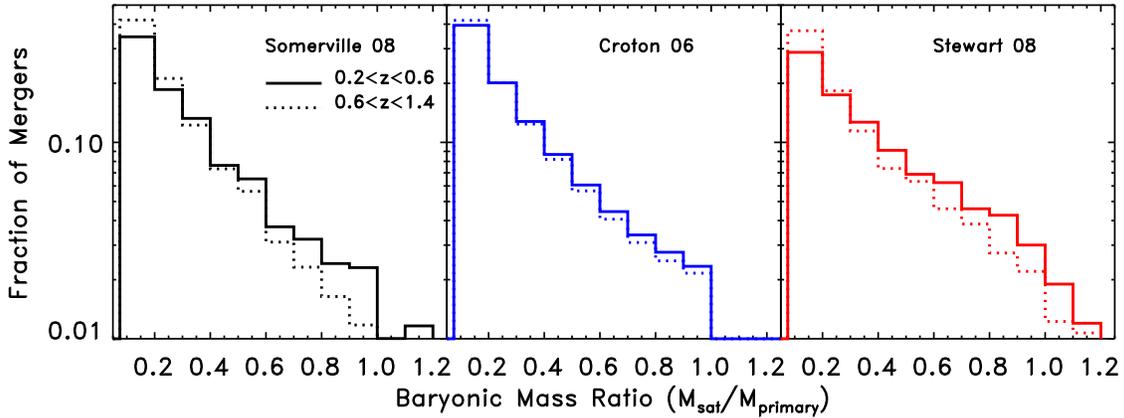}
\caption{The normalized distribution of baryonic mass ratios
for galaxy mergers selected from the Somerville et al. 2008 (black
histogram), Croton et al. 2006 (blue), and Stewart et al. 2009b (red)
models. The mergers were selected to be in two redshift bins ( $0.2 < z <0.6$; solid
lines; $0.6 < z < 1.4$ dashed lines), 
with total stellar masses ($M_{star,1} + M_{star,2}) > 10^{10} M_{\odot}$), and
baryonic mass ratios $\geq$ 1:10. There is good agreement between all
three models, and little evolution in the distribution with redshift. (Baryonic mass ratios greater than unity arise when the
more massive galaxy has less baryons but more dark matter than
its companion).}
\end{figure*}

\begin{figure}
\plotone{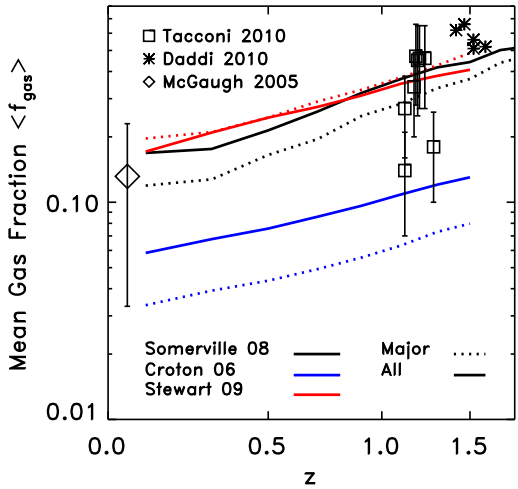}
\caption{ The predicted mean baryonic gas fraction $\langle f_{gas} \rangle$ v.
redshift for 1:1 - 1:10 baryonic mass ratio mergers (solid) and major 
1:1-1:4 baryonic mass ratio mergers (dashed line). All models
predict strong evolution in $\langle f_{gas} \rangle$  with redshift, but the Somerville
et al. 2008 (black lines) and Stewart et al. 2009b (red lines) mergers
have higher $\langle f_{gas} \rangle $ than the Croton et al. 2006 models (blue lines).
For comparison, we have plotted the observed $f_{gas}$ at $z > 1$ for
the Tacconi et al. 2010 (squares) and Daddi et al 2010 (asterisks)
samples, and the average $f_{gas}$ and standard deviation for the McGaugh 2005
sample of local $M_{star} > 10^{10} M_{\odot}$ disk galaxies (diamond).}
\end{figure}

\subsection{Predicted Distribution of Merger Properties}
We calculate the cosmologically-averaged observability
timescale $\langle T_{obs}(z) \rangle$ for each method for finding mergers
from the individual $T_{obs}$ given in Figures 3 and 4 and assumptions 
about the distribution of galaxy merger mass ratios and $f_{gas}$ as follows:
\begin{equation}
\langle T_{obs}(z) \rangle = \sum_{i,j}  w_{i,j}(z) \times T_{i,j} 
\end{equation}
where $w_{i,j}(z)$ is the fraction of mergers at redshift z with baryonic 
mass ratio $i$ and baryonic gas fraction $j$, and $T_{i,j}$
is the observability timescale for a merger with baryonic
mass ratio $i$ and baryonic gas fraction $j$.  
(Although the timescales do not change much with progenitor mass, in practice
 we also sum over two bins in primary progenitor baryonic mass based on the
input simulation primary galaxy masses of $ 2.0 \times 10^{10}$, $6.2 \times 10^{10} M_{\odot}$). 

We currently have very little empirical knowledge of
the distribution of baryonic mass ratios or gas fractions
for merging galaxies. Therefore we assume the relative 
distributions of mass ratios, gas fractions, and mass
as a function of redshift predicted by three different
cosmological-scale galaxy evolution models: Somerville
et al. 2008 [S08]; Croton et al. 2006 [C06]; and Stewart et al. 
2009b [St09]. Note that for the timescale calculations, we make use of only
the {\it relative} distributions of merger properties from
these simulations and therefore are independent of their predicted galaxy merger rates.

Each of these models starts with a mass distribution
and merger history (“merger tree”) for the dark matter
central and sub-halos. S08 uses an analytically-derived
dark matter merger tree (Somerville \& Kolatt 1999)
 with a Sheth \& Tormen (1999) dark matter halo
mass function; the C06 and St09
merger trees are derived from N-body simulations (the
Millennium Simulation from Springel et al. 2005 and an
Adaptive Refinement Tree N-body simulation from A.
Klypin described in Stewart et al. 2008, respectively). 
These dark matter halos are populated with
 galaxies, either via a semi-analytic approach
that adopts physical prescriptions for galaxy formation (C06, S08)
or via a semi-empirical approach that matches the expected 
clustering and abundances of dark matter halos
to the observed galaxy population (St09). In the semi-analytical models, 
gas cools onto the dark matter halos and
is converted into stars assuming the Kennicutt-Schmidt
law (Kennicutt et al. 1998). The semi-analytic models also
adopt similar prescriptions for feedback from supernovae
and both high-luminosity quasars and lower luminosity
active galactic nuclei.  The S08 models are tuned to match
the gas fractions of local galaxies.  By contrast, the semi-empirical model of 
St09 assigns the stellar masses and gas masses of galaxies
based on observationally-derived correlations, with stellar masses following the abundance matching
technique of Conroy \& Wechsler (2009), and cold gas fractions
 following an empirical fit to the results of McGaugh (2005) at 
$z \sim 0$  and Erb et al. (2006) at $z \sim 2$ (see Stewart et al. 2009a,b 
for additional details).

\begin{figure*}
\plotone{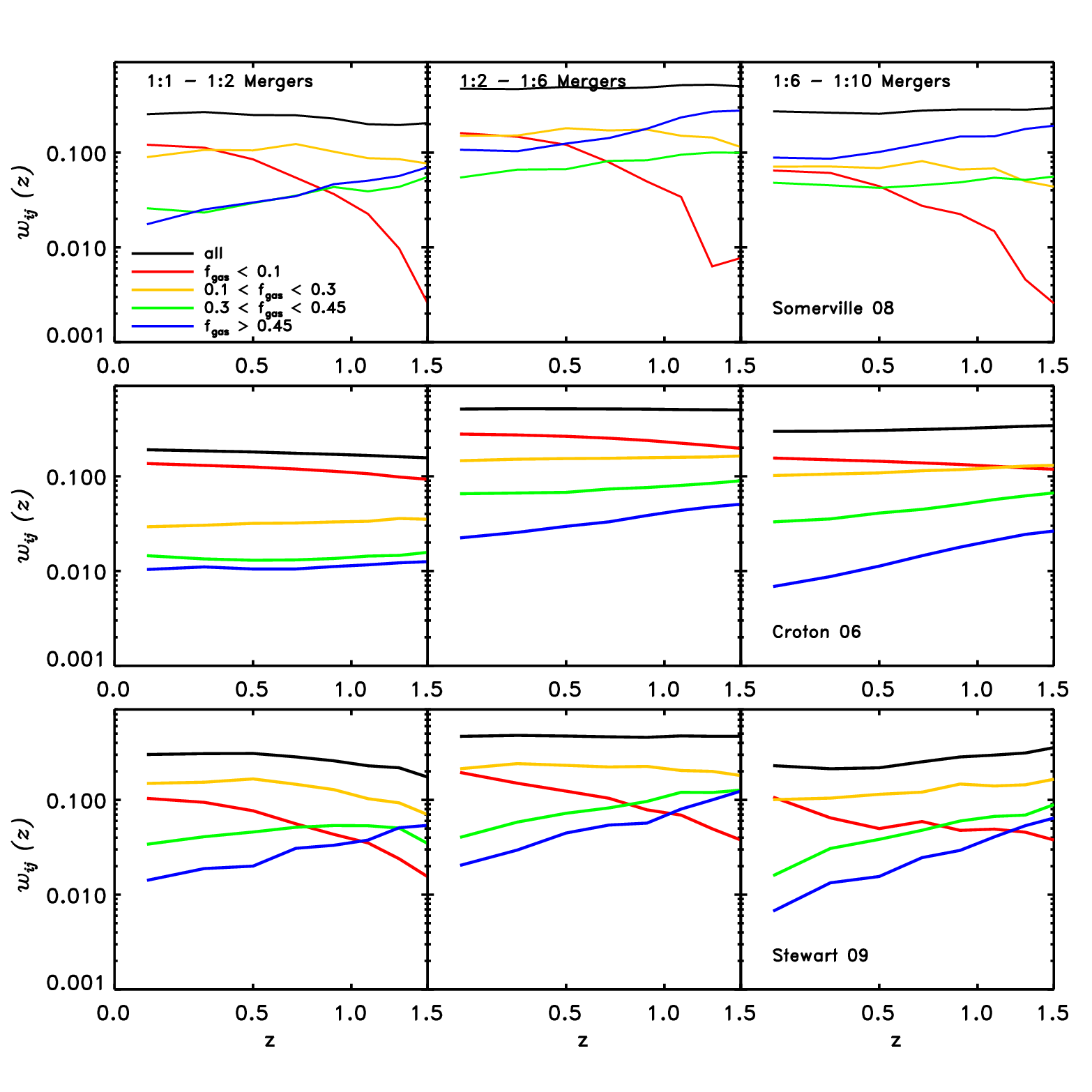}
\caption{The fraction of mergers $w_{i,j}$ with baryonic mass ratio $i$ and 
baryonic gas fraction $j$ v. redshift from the Somerville et al. (2008)
models (top panels), the Croton et al. (2006) models (middle panels), and 
Stewart et al. (2009b) models (bottom panels). The weights
have been calculated for three baryonic mass ratio bins ( 1:1-1:2, left panels; 
1:2-1:6, center panels; 1:6-1:10, right panels) and four baryonic gas
fraction bins ($f_{gas} < 0.1$, red lines; $0.1 < f_{gas} < 0.3$, orange lines; 
$0.3 < f_{gas} < 0.45$, green lines; $f_{gas} > 0.45$, blue lines.) The black lines
in each panel show the fraction of mergers of all gas fractions for that panel's mass ratio bin.
}
\end{figure*}

For each galaxy evolution model, we select galaxy
mergers with total stellar masses ($M_{star,1} + M_{star,2}$) 
$\ge 10^{10} M_{\odot}$, baryonic mass ratios $\ge$ 1:10, and 
$0 < z < 1.5$.     The galaxy stellar mass functions at $z < 1.5$ from 
all of the models are in good agreement with each other.  
St09 adopts the sub-halo mass - stellar mass matching relationship
from Conroy \& Wechsler (2009),  based upon 
the observed stellar mass functions at $0 < z < 2$ (Bell et al. 2003, 
Panter et al. 2007,  Dory et al. 2005, Borsch et al. 2006, P\'{e}rez-Gonz\'{a}lez et al. 2008, 
Fontana et al. 2006).   The S08 and C06 predictions for the galaxy stellar mass functions
(assuming a Chabrier initial mass function and Bruzual \& Charlot 2003 SEDs) 
are found to be in good agreement with these same observed measurements at $z < 1.5$
and for galaxies with masses above $\sim 10^{10} M_{\odot}$ 
(see  Fontanot et al. 2009,  Kitzbichler \& White 2007). 

We compare the distribution of baryonic mass ratios and
gas fractions for the three models in Figures 5 and 6.
In Figure 5, we find that the S08, St09, and C06 models predict similar baryonic mass ratio distributions
despite the different derivations of the dark matter halo mass function, merger tree,
and baryonic masses.   We find a small increase in the relative fraction of 
minor mergers ( $M_{sat} / M_{primary} < 0.25$) at higher redshifts for the
S08 and St09 models, while C06 predicts no change in the mass ratio distribution. 

However, the models make different predictions about
the merger gas properties (Figure 6). In all the models,
$f_{gas}$ increases significantly with redshift such that the
mean baryonic gas fraction of the mergers roughly doubles 
from $z = 0$ to $z = 1$. The mean $f_{gas}$ values for all 1:1
- 1:10 mergers predicted by the S08 and St09 models are
in good agreement. On the other hand, the mean $f_{gas}$ in
the C06 models is much lower at all redshifts. Both S08 and
C06 predict that major mergers with baryonic mass ratios less than 1:4 
have lower mean gas fractions than
minor mergers. On the other hand, St09 predict that
major mergers have $f_{gas}$ similar to the overall merger
population. 

In Figure 6, we also compare the predicted mean gas fractions 
to recent measurements of the cold gas fraction for
massive disk galaxies at $z \sim  1- 1.5$ from Tacconi et al.
(2010) and Daddi et al. (2010). Although these pioneering 
studies have small and biased samples, they indicate
that the typical baryonic gas fractions of massive disk
galaxies are several times higher at $z \sim 1 - 1.5$ than
locally (e.g McGaugh et al. 2005), in reasonable agreement 
with the Somerville et al. (2008) and Stewart et al. (2009b) models.

For each model, we compute the weight $w_{i,j}(z)$ for
three bins in baryonic mass ratio and four bins in $f_{gas}$
for $\delta z = 0.2$ redshift bins from $z = 0$ to $z = 1.6$, 
normalized to all merging systems in each redshift bin with
($M_{star,1} +M_{star,2}$) $\ge 10^{10} M_{\odot}$ and baryonic mass ratios
$\ge$ 1:10 (Figure 7). The spacing of these bins is determined 
by the simulation parameter space for the SDSS-motivated 
galaxy mergers: 1:1 - 1:2 (major), 1:2-1:6
(intermediate), and 1:6-1:10 (minor) baryonic mass ratio bins; 
and 0.0-0.1, 0.1-0.3, 0.3-0.45, and 0.45-1.0 $f_{gas}$
bins.  The three baryonic mass ratio bins are plotted in separate panels 
(major:left, intermediate:center, and minor:
right). The four $f_{gas}$ bins are plotted with different color
lines in each mass ratio panel. The resulting weights
are shown for the three galaxy evolution models (S08:
top, C06: middle, St09: bottom). 

For all three models, the baryonic mass ratio distribution changes 
little with redshift, and intermediate mass
ratio mergers (1:2 - 1:6; middle panels) are ∼ 50\% of the
‘total’ (1:1 -1:10) merger population at all redshifts. The
predicted weights $w_{i,j}(z)$ vary most in the predicted distribution 
and evolution of merger gas fractions. All three
models predict that lower gas fraction mergers dominate
the $z = 0$ merger population (red and yellow lines). However, 
S08 predicts strong evolution in the gas properties
of mergers, such that by $z \ge 1$,  the merger population
is dominated by high gas fraction ($f_{gas} > 0.45$) intermediate 
and minor mergers (blue lines, middle and right
panels). Because C06 predicts low mean $f_{gas}$ at $z < 1.5$, 
they predict that the lowest gas fraction mergers continue to dominate at $z > 1$ 
(red lines). Like S08, the St09 model predicts higher mean $f_{gas}$ at $z \ge 1$. But
St09 has a different distribution of $f_{gas}$ at a given
redshift. St09 has fewer very gas-rich mergers and more
very gas-poor mergers at $z \sim 1$, and predicts that most
mergers at $0 < z < 1.5$ have intermediate $f_{gas}$( $\sim 0.2$)
in all mass ratio bins (yellow lines). As we will discuss
in the next section, the relative frequency of gas-rich intermediate 
and minor mergers at $z \ge 1$ has important
implications for the interpretation of asymmetric galaxies at this epoch.

\begin{figure*}
\plotone{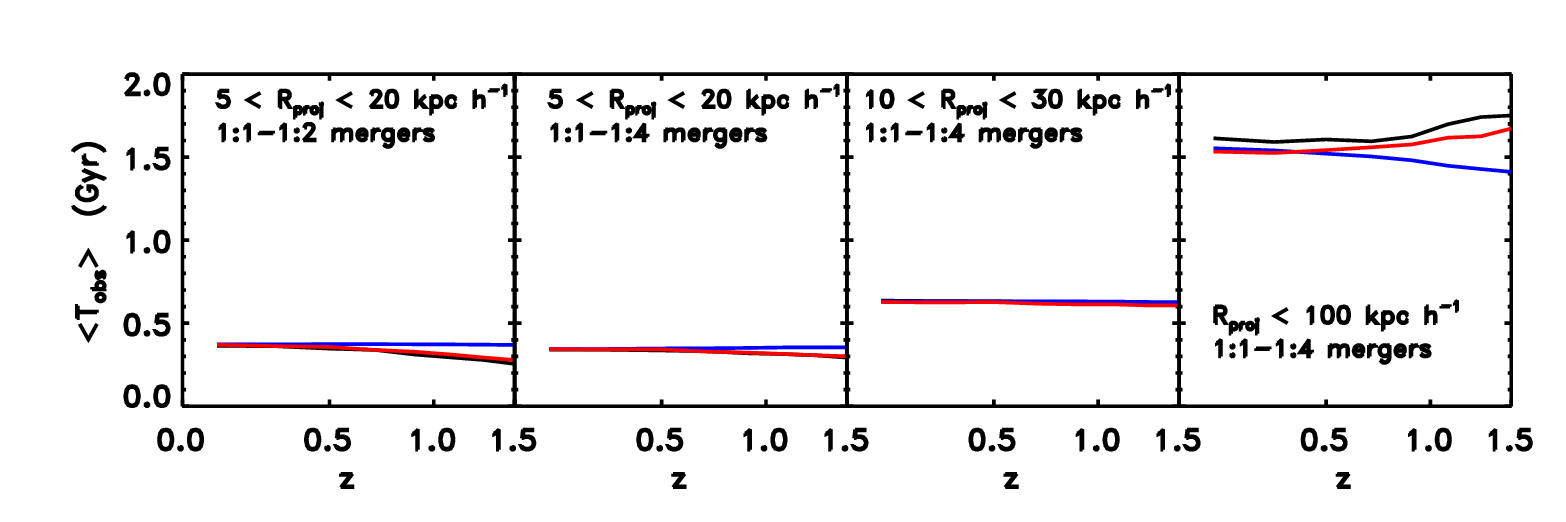}
\caption{The average observability timescale $\langle T_{obs} \rangle$ 
for several close pair selection criteria: $5 < R_{proj} < 20$ kpc $h^{-1}$ and 1:1 - 1:2
baryonic mass ratio; $5 < R_{proj} < 20$ kpc $h^{-1}$ and 1:1 - 1:4 baryonic mass 
ratio; $10 < R_{proj} < 30$ kpc $h^{-1}$ and 1:1 - 1:4 baryonic mass
ratio; $R_{proj} < 100$ kpc $h^{-1}$ and 1:1 - 1:4 baryonic mass ratio. 
The close pair timescales are very similar for all these models. }
\end{figure*}

\begin{figure}
\plotone{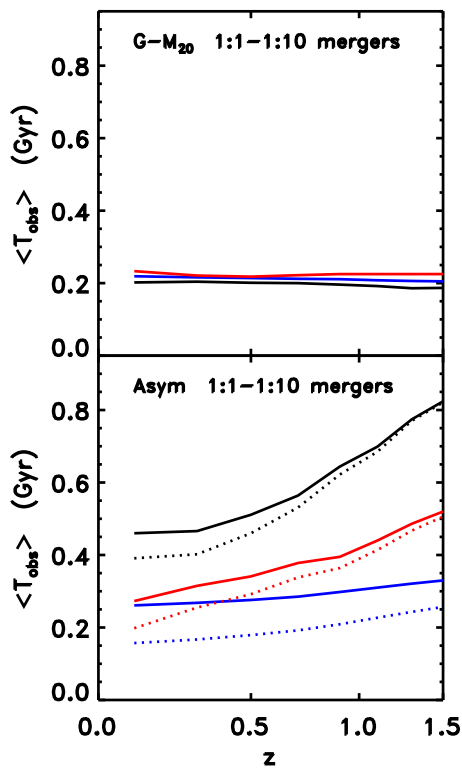}
\caption{The average observability timescale $\langle T_{obs} \rangle$ 
for $G-M_{20}$-selected mergers (top) and asymmetric mergers (bottom) 
with baryonic mass ratios between 1:1 and 1:10.  $G - M_{20}$ timescales are
independent of the assumptions about the distribution of $f_{gas}$.  Asymmetry
timescales evolve most strongly for the S08 model $f_{gas}$ distribution (black
lines), which assumes many gas-rich minor mergers at $z \ge 1$.  Asymmetry
timescales evolve weakly for the C06 model $f_{gas}$ distribution (blue lines),
which assumes low $f_{gas}$ mergers at $0 < z < 1.5$. The $A$ timescales
predicted by the St09 models also evolve strongly with redshift, but
are shorter than S08 because St09 has fewer very gas-rich mergers.
The dotted lines assume that $f_{gas} < 0.1$ mergers are not detected
by $A$ (see text). }
\end{figure}

\subsection{ Average Merger Timescales $\langle T_{obs}(z) \rangle$ }
We compute $\langle T_{obs}(z) \rangle$ for $G - M_{20}$, asymmetry, and
close pairs with $5 < R_{proj} < 20$, and $10 < R_{proj} < 30$ kpc $h^{-1}$, 
using Eqn. 8,  the $T_{i,j}$ computed from the
{\sc GADGET-SUNRISE} simulations as a function of baryonic mass 
ratio and gas fraction (Figures 1 and 2), and the relative weights $w_{i,j}$ 
computed from each of the cosmological galaxy evolution models.
These cosmologically-weighted timescales $\langle T_{obs}(z) \rangle$ are
plotted as a function of redshift for the three different
models for each method for finding mergers (Figures 8
and 9).

The relative weights $w_{i,j}$ are adjusted for the mass ratio 
sensitivity for each technique, so that the resulting
merger rates are not extrapolated to mass ratios below
what is detected. Therefore the weights for $G - M_{20}$
and $A$  are normalized for all mergers with baryonic 
mass ratios between 1:1 and 1:10, and assume no
mergers with mass ratios $\leq$ 1:10 are detected by these
methods (same as Fig. 7). The weights for close pairs
are normalized to baryonic mass ratios between 1:1 and
1:4 or between 1:1 and 1:2, depending on the stellar mass
and luminosity ratios adopted by the studies presented
here (Table 1).

Both the gas fraction and morphological make-up of merger
progenitors are expected to change significantly between 
$0 < z < 1.5$.    The number density of red sequence (presumably
gas-poor) galaxies more massive than $\sim 10^{10} M_{\odot}$ roughly
doubles over this time period (Bell et al. 2004,  Faber et al. 2007), and 
the fraction of bright galaxies which are spheroid dominated increases 
from $\sim$ 20\%  at $z \sim 1.1$ to $\sim$ 40\% at $z \sim 0.3$
(Lotz et al. 2008a).    The cold  gas fractions measured for
small numbers of disk galaxies at $z \sim 1 - 1.5$ are $\sim$ 40\%, 
three times higher than today's $\sim$ 10\%  (Tacconi et al. 2010; 
Daddi et al. 2010;  McGaugh 2005). 

For $G-M_{20}$ and close pairs,  the individual observability timescales
 $T_{i,j}$ are not a strong function of $f_{gas}$. We
assume that $f_{gas} < 0.1$ mergers have the same observability 
timescales as $f_{gas} = 0.2$ mergers of the same mass
ratio.    We note that gas-free spheroid-spheroid merger simulations 
presented in Bell et al. (2005) imply $\sim$0.2 Gyr timescales for visual
classification of double nuclei,  consistent with our adopted $G-M_{20}$
timescale for $f_{gas} < 0.1$. 

 However, the $A$ timescales are a strong function
of $f_{gas}$. Therefore, we compute upper and lower limits
to $\langle T_{obs}(z) \rangle$ by assuming that that $f_{gas} < 0.1$ 
mergers have the same observability timescales as $f_{gas} = 0.2$
mergers (Figure 9; upper limit, solid lines) and by assuming that 
$f_{gas} < 0.1$ mergers are not detected by $A$ (Figure 9: lower limit, dotted lines).
We note that  Bell et al. (2006) found visual-classification timescales 
for gas-poor spheroid-spheroid major merger simulations of $T_{obs} \sim 0.15$ Gyr, 
intermediate between the $T_{obs} = 0 $  and $T_{obs} \sim 0.3$ Gyr 
for $f_{gas} = 0$ adopted here.  

The cosmologically-averaged timescales for close pairs
do not evolve with redshift and do not depend on the
adopted cosmological galaxy evolution model (Fig. 8).
This is not surprising given that the close pair timescales
do not depend on gas fraction (which evolves strongly with
redshift). The individual close pair timescales do depend
on mass ratio, but the relative distribution of mass ratios
is not predicted to evolve with redshift. The timescales
for close pairs selected with baryonic mass ratios between
1:1 and 1:4 are $\sim$ 0.33 Gyr for $5 < R_{proj} < 20$ kpc $h^{-1}$,
and $\sim$ 0.63 Gyr for $10 < R_{proj} < 30$ kpc $h^{-1}$ (Table 1;
Figure 8).

\begin{figure*}
\plotone{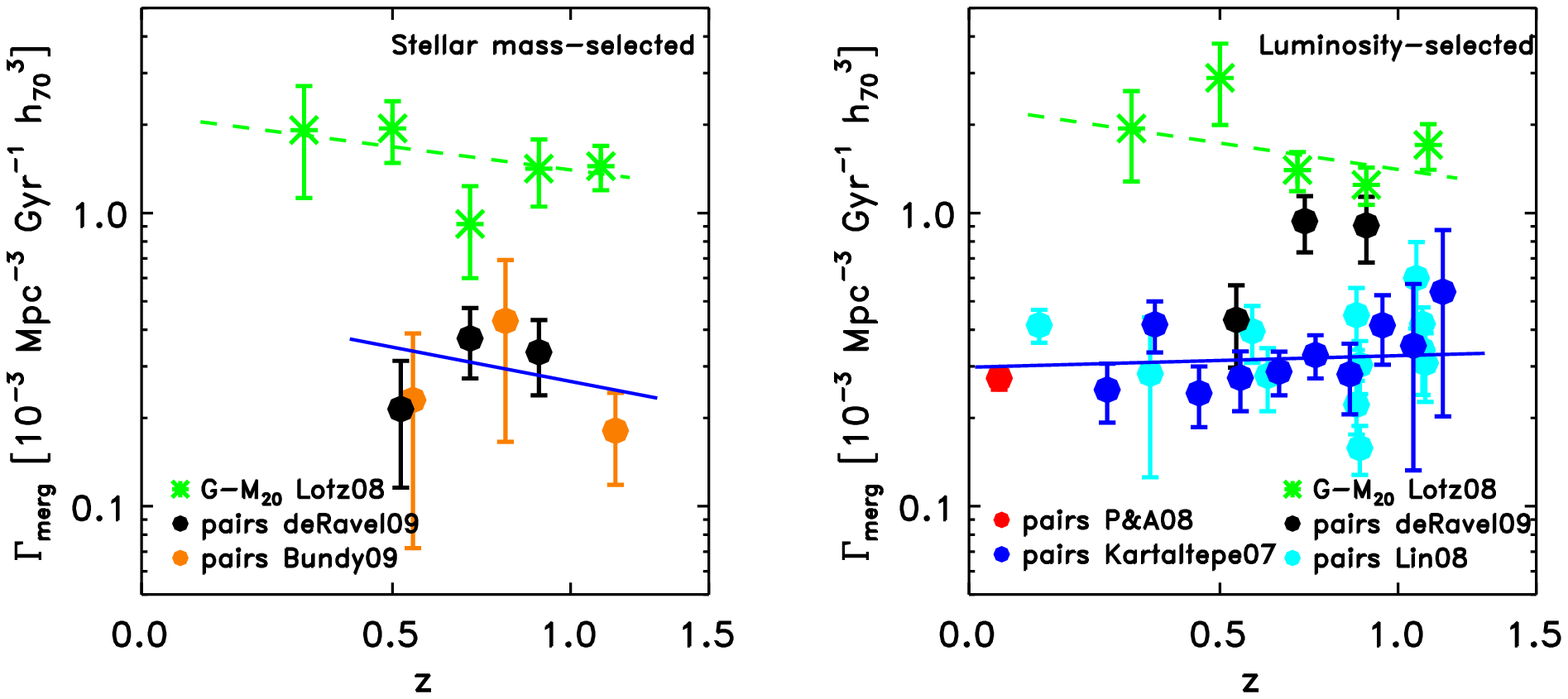}
\plotone{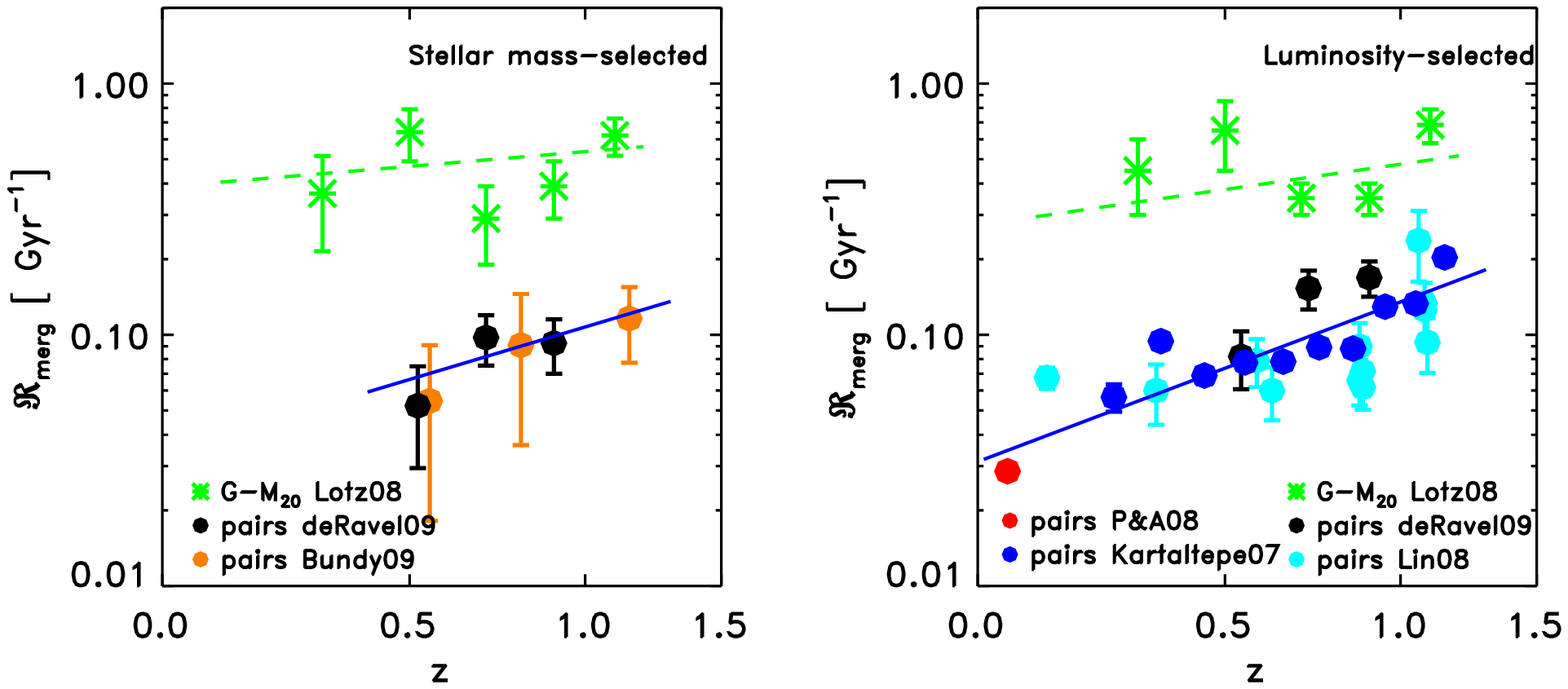}
\caption{Top: $\Gamma_{merg}$, the merger rate per co-moving unit volume, for close pairs (circles) and $G-M_{20}$ (asterisks), 
for stellar mass-selected (left) and rest-frame luminosity selected samples. Bottom: $\Re_{merg}$, the fractional merger rate,  for close pairs (circles) and
$G-M_{20}$ (asterisks), for the same samples. The error-bars are computed using the observational uncertainties on $f_{merg}$, $f_{pair}$,  and $n_{gal}$ and do not include uncertainties in $\langle T_{obs} \rangle$.  $G-M_{20}$ probes both major and minor mergers, and therefore captures a `total' merger rate, which is several times higher than the major merger rate probed by these close pair studies. 
The evolution in $\Gamma_{pairs}(z)$ is weaker than in $\Re_{pairs}(z)$ because $f_{pairs}$ increases
with redshift  (Fig. 1) while the corresponding $n_{gal}$ decreases with redshift for fixed stellar mass and PLE galaxy selections (Fig. 2).  The best-fit slopes for the close pair (major) merger rates (blue solid lines) are given in \S5.1 and the best-slopes for the $G-M_{20}$ (total) 
merger rates (green dashed lines) are given in \S5.2. }
\end{figure*}

We find that $\langle T_{obs}(z) \rangle$ for $G-M_{20}$ 
is also very similar for all three model weights, and does not evolve with
redshift. This is not surprising as $G-M_{20}$ $T_{obs}$ does not
correlate with $f_{gas}$ nor baryonic mass ratio between 1:1
and 1:10. Therefore, the $G-M_{20}$ observability timescales
are not sensitive to the underlying assumptions about the
distribution of merger gas fractions or mass ratios. For
mergers with stellar masses $\ge 10^{10} M_{\odot}$ at $0 < z < 1.5$,
$\langle T_{obs}(z) \rangle$ is $\sim$ 0.21 Gyr for $G-M_{20}$ 
(Table 2; Figure 9 upper panel).

The cosmologically-averaged timescales for asymmetry
are the most sensitive to the assumptions about the joint
distribution of merger gas fractions and mass ratios (Figure 9). 
The uncertainty in the asymmetry timescales of
$f_{gas} < 0.1$ gas-poor mergers changes $\langle T_{obs}(A) \rangle$ by less
than 0.1 Gyr (dotted lines), and therefore is not a major
contributor to the uncertainty in $\langle T_{obs}(A,z) \rangle$. 
However, the  $\langle T_{obs}(A,z) \rangle$ calculated with the S08 weights (black
lines) are significantly higher than those calculated with
the C06 (blue lines) and St09 (red lines) weights. This
arises from the strong dependence of $A$ timescales on
the $f_{gas}$ and the different predictions for the distribution
of $f_{gas}(z)$ for the different models. The strong evolution of 
the mean $f_{gas}$ and higher frequency of very gas-rich minor
 mergers predicted by S08 result in the strong
evolution of $\langle T_{obs}(A,z) \rangle$ with redshift. The low mean
$f_{gas}$ and its weak evolution predicted by C06 results
in short $\langle T_{obs}(A,z) \rangle$ that evolves weakly with redshift.
The St09 models predict similar evolution in the mean
$f_{gas}$ to S08, but has fewer high gas fraction 
intermediate/minor mergers, and therefore predicts lower values
for $\langle T_{obs}(A,z) \rangle$.

In summary, the average observability timescales for
$G-M_{20}$ and close pairs are not expected to evolve with
redshift between $0 < z < 1.5$, and are relatively insensitive
 to the distribution of merger properties. The average
observability timescales for $A$, on the other hand, are expected
 to increase between $0 < z < 1.5$ as the mean gas
fraction of mergers increase, and are highly sensitive to
both the distribution of merger gas fractions and mass
ratios.  If the typical gas fractions of galaxy mergers at
$z \ge  1$ are as high as 0.4, as suggested by recent molecular
gas observations, then typical timescales for identifying a
merger as asymmetric should more than double. If very
gas-rich minor mergers are also more likely at $z \ge 1$,
than the contribution of minor mergers to the asymmetric 
galaxy population will also increase with redshift.

\section{GALAXY MERGER RATES}
In the previous section, we calculated cosmologically-averaged
merger observability timescales that account for differences in 
merger identification techniques and the distribution of merger mass 
ratio and gas fraction. In this section, we use these improved estimates
of $\langle T_{obs}(z) \rangle$ to re-analyze recent studies of the evolution 
of the galaxy mergers (reviewed in \S3). In Tables 1, 2, and 3, we give
$\langle T_{obs}(z) \rangle$ calculated specifically for each merger study/technique, 
the resulting fractional merger rates, $\Re_{merg}(z)$, and the merger rates
per co-moving volume,  $\Gamma_{merg}(z)$. For close pairs and
$G-M_{20}$, we adopt the S08 timescales, given that there
is little variation in timescales with cosmological merger
distributions. For asymmetry, different gas fraction assumptions 
result in very different $\langle T_{obs}(z) \rangle$. There are
also very different literature values for the asymmetry
fractions, hence we treat each set of observations/model
timescales separately.

We calculate both $\Gamma_{merg}(z)$ and $\Re_{merg}(z)$ for close
pairs, $G-M_{20}$, and $A$ selected mergers. We focus primarily 
on the galaxy merger rates for samples selected with
an evolving luminosity cut (‘PLE’) or fixed stellar mass
cut (‘$M_{star}$’) in \S5.1-5.3. In \S5.4 we discuss the implications 
of selecting parent galaxy samples with a constant
number density.   In \S5.5 we compare our results for stellar-mass
selected merger rates to the predictions of several galaxy evolution models. 
In general, we fit the galaxy merger rates with  power-laws of
the form $C \times (1+z)^{\alpha}$.

\begin{deluxetable*}{lcccccccc}
\tablewidth{0pc}
\tablecolumns{9}
\tabletypesize{\scriptsize}
\tablecaption{Close Pair Fractions and Merger Rates}
\tablehead{ \colhead{$z$} & \colhead{$N_c$}  & \colhead{$f_{pair}$}   & \colhead{$n_{gal}$ }  & \colhead{$\langle T_{obs} \rangle_{S08}$}  & \colhead{$\langle T_{obs} \rangle_{C06}$}  
& \colhead{$\langle T_{obs} \rangle_{St09}$}  &\colhead{$\Re_{merg,S08}$\tablenotemark{a}} & \colhead{$\Gamma_{merg, S08}$\tablenotemark{a}}  \\
 \colhead{}   & \colhead{}   &\colhead{}   &\colhead{[$10^{-3}$ Mpc$^{-3}$]}  &\colhead{ [Gyr]}  &\colhead{ [Gyr]} &\colhead{ [Gyr]} & \colhead{ [Gyr$^{-1}$]} 
 &\colhead{  [$10^{-3}$ Gyr$^{-1}$ Mpc$^{-3}$]}}
\startdata
\cutinhead{Luminosity-selected Pairs}
\hline
\sidehead{Patton \& Atfield 2008; $5 < R_{proj} < 20 $ kpc $h^{-1}$;  $1 < L_{primary}/L_{sat} < 2$;  $\Delta V < 500$ km s$^{-1}$; $ -22 < M_r < -18 $ } 
0.05 &  0.021 $\pm$ 0.001 & \nodata & 9.55 $\pm$ 0.07 & 0.36 & 0.37 & 0.37  & 0.0175 $\pm$ 0.0008 &  0.17 $\pm$ 0.01 \\
\hline 
\sidehead{Lin et al. 2008; $10 < R_{proj} < 30$ kpc $h^{-1}$; $1 < L_{primary}/L_{satellite} < 4$; $\Delta V < 500$ km s$^{-1}$; $-21 < M_B + 1.3z < -19$}
0.12 & 0.054 $\pm$ 0.005 & 0.072 $\pm$ 0.007 & 6.14 $\pm$ 0.52 & 0.64 & 0.63 & 0.63 & 0.068 $\pm$ 0.007 & 0.41 $\pm$ 0.05\\
0.34 & 0.041 $\pm$ 0.011 & 0.063 $\pm$ 0.017 & 4.72 $\pm$ 2.30 & 0.63 & 0.63 & 0.63 & 0.060 $\pm$ 0.016 & 0.28 $\pm$ 0.16\\
0.58 & 0.063 $\pm$ 0.014 & 0.083 $\pm$ 0.018 & 5.00 $\pm$ 0.16 & 0.63 & 0.63 & 0.63 & 0.079 $\pm$ 0.017 & 0.40 $\pm$ 0.09\\
0.62 & 0.047 $\pm$ 0.011 & 0.063 $\pm$ 0.015 & 4.64 $\pm$ 0.21 & 0.63 & 0.63 & 0.63 & 0.060 $\pm$ 0.014 & 0.28 $\pm$ 0.07\\
0.87 & 0.044 $\pm$ 0.009 & 0.068 $\pm$ 0.014 & 3.37 $\pm$ 0.10 & 0.62 & 0.63 & 0.61 & 0.066 $\pm$ 0.014 & 0.22 $\pm$ 0.05\\
0.87 & 0.060 $\pm$ 0.014 & 0.093 $\pm$ 0.022 & 4.98 $\pm$ 0.15 & 0.62 & 0.63 & 0.61 & 0.090 $\pm$ 0.021 & 0.45 $\pm$ 0.11\\
0.88 & 0.048 $\pm$ 0.010 & 0.074 $\pm$ 0.015 & 4.24 $\pm$ 0.13 & 0.62 & 0.63 & 0.61 & 0.072 $\pm$ 0.015 & 0.30 $\pm$ 0.06\\
0.88 & 0.041 $\pm$ 0.008 & 0.064 $\pm$ 0.012 & 2.55 $\pm$ 0.08 & 0.62 & 0.63 & 0.61 & 0.062 $\pm$ 0.012 & 0.16 $\pm$ 0.03\\
1.06 & 0.159 $\pm$ 0.050 & 0.249 $\pm$ 0.078 & 2.53 $\pm$ 0.24 & 0.63 & 0.63 & 0.61 & 0.237 $\pm$ 0.074 & 0.60 $\pm$ 0.20\\
1.08 & 0.083 $\pm$ 0.023 & 0.132 $\pm$ 0.037 & 2.71 $\pm$ 0.26 & 0.63 & 0.63 & 0.61 & 0.126 $\pm$ 0.035 & 0.34 $\pm$ 0.10\\
1.08 & 0.088 $\pm$ 0.009 & 0.140 $\pm$ 0.014 & 3.14 $\pm$ 0.30 & 0.63 & 0.63 & 0.61 & 0.133 $\pm$ 0.013 & 0.42 $\pm$ 0.06\\
1.09 & 0.061 $\pm$ 0.015 & 0.098 $\pm$ 0.024 & 3.30 $\pm$ 0.32 & 0.63 & 0.63 & 0.61 & 0.093 $\pm$ 0.023 & 0.31 $\pm$ 0.08\\
\hline
\sidehead{de Ravel et al. 2009; $R_{proj} < 100$ kpc h$^{-1}$; $1 < L_{primary}/L_{satellite} < 4$; $\delta V < 500$ km s$^{-1}$; $M_B < -18.77 - 1.11z$}
0.54 & \nodata & 0.22 $\pm$ 0.06 & 5.27$^{+0.92}_{-0.86}$ &  1.61 & 1.52 & 1.54 & 0.082 $\pm$ 0.021 & 0.43 $\pm$ 0.13 \\
0.72 & \nodata & 0.41 $\pm$ 0.07 & 6.13$^{+0.78}_{-0.76}$ &  1.60 & 1.50 & 1.56 & 0.153 $\pm$ 0.027 & 0.94 $\pm$ 0.20\\
0.90 & \nodata & 0.46 $\pm$ 0.07 & 5.37$^{+1.05}_{-0.99}$ &  1.62 & 1.48 & 1.58 & 0.169 $\pm$ 0.027 & 0.91 $\pm$ 0.23\\
\hline
\sidehead{Kartaltepe et al. 2007; $5 < R_{proj} < 20$ kpc $h^{-1}$; $\delta z < 0.05$; $M_V < -19.8 - 1.0z$}
0.25 & \nodata & 0.032 $\pm$ 0.004 & 4.42$^{+0.84}_{-0.78}$ & 0.34 & 0.34 & 0.35 &0.056 $\pm$ 0.007 &0.25$^{+0.06}_{-0.05}$  \\
0.35 & \nodata & 0.055 $\pm$ 0.003 & 4.42$^{+0.84}_{-0.78}$ & 0.35 & 0.35 & 0.34 &0.094 $\pm$ 0.005 &0.42 $\pm$ 0.08\\
0.45 & \nodata & 0.039 $\pm$ 0.003 & 3.53$^{+0.78}_{-0.72}$ & 0.34 & 0.35 & 0.34 &0.069 $\pm$ 0.005 &0.24$^{+0.06}_{-0.05}$\\
0.55 & \nodata & 0.044 $\pm$ 0.003 & 3.53$^{+0.78}_{-0.72}$ & 0.34 & 0.35 & 0.34 &0.078 $\pm$ 0.005 &0.27$\pm$0.06\\
0.65 & \nodata & 0.043 $\pm$ 0.002 & 3.68$^{+0.60}_{-0.56}$ & 0.33 & 0.35 & 0.33 &0.078 $\pm$ 0.004 &0.29$\pm$0.05\\
0.75 & \nodata & 0.049 $\pm$ 0.002 & 3.68$^{+0.60}_{-0.56}$ & 0.33 & 0.35 & 0.33 &0.089 $\pm$ 0.004 &0.33$^{+0.06}_{-0.05}$\\
0.85 & \nodata & 0.047 $\pm$ 0.003 & 3.20$^{+0.84}_{-0.75}$ & 0.32 & 0.35 & 0.33 &0.088 $\pm$ 0.006 &0.28$^{+0.08}_{-0.07}$\\
0.95 & \nodata & 0.069 $\pm$ 0.003 & 3.20$^{+0.84}_{-0.75}$ & 0.32 & 0.35 & 0.32 &0.129 $\pm$ 0.006 &0.41$^{+0.11}_{-0.10}$\\
1.05 & \nodata & 0.071 $\pm$ 0.003 & 2.65$^{+1.65}_{-2.04}$ & 0.32 & 0.35 & 0.32 &0.133 $\pm$ 0.006 &0.35$^{+0.22}_{-0.27}$\\
1.15 & \nodata & 0.105 $\pm$ 0.004 & 2.65$^{+1.65}_{-2.04}$ & 0.31 & 0.35 & 0.31 &0.203 $\pm$ 0.008 &0.54$^{+0.34}_{-0.42}$\\
\hline
\sidehead{Kartaltepe et al. 2007; $5 < R_{proj} < 20$ kpc $h^{-1}$; $\delta z < 0.05$; $M_V < -19.8$ }
0.15 & \nodata & 0.021 $\pm$ 0.007 & 5.336$^{+0.941}_{-0.828}$   & 0.34 &0.34 &0.35 &0.04 $\pm$ 0.01 &0.20$^{+0.07}_{-0.07}$\\
0.25 & \nodata & 0.018 $\pm$ 0.004 & 5.543$^{+1.049}_{-0.980}$   & 0.34 &0.34 &0.35 &0.03 $\pm$ 0.01 &0.18$^{+0.05}_{-0.05}$\\
0.35 & \nodata & 0.040 $\pm$ 0.004 & 5.543$^{+1.049}_{-0.980}$   & 0.35 &0.35 &0.34 &0.07 $\pm$ 0.01 &0.38$^{+0.08}_{-0.08}$\\
0.45 & \nodata & 0.030 $\pm$ 0.003 & 5.040$^{+1.111}_{-1.024}$   & 0.34 &0.35 &0.34 &0.05 $\pm$ 0.01 &0.27$^{+0.06}_{-0.06}$\\
0.55 & \nodata & 0.034 $\pm$ 0.003 & 5.040$^{+1.111}_{-1.024}$   & 0.34 &0.35 &0.34 &0.06 $\pm$ 0.01 &0.30$^{+0.07}_{-0.07}$\\
0.65 & \nodata & 0.033 $\pm$ 0.002 & 6.321$^{+1.035}_{-0.966}$   & 0.33 &0.35 &0.33 &0.06 $\pm$ 0.00 &0.38$^{+0.07}_{-0.06}$\\
0.75 & \nodata & 0.044 $\pm$ 0.002 & 6.321$^{+1.035}_{-0.966}$   & 0.33 &0.35 &0.33 &0.08 $\pm$ 0.00 &0.51$^{+0.09}_{-0.08}$\\
0.85 & \nodata & 0.045 $\pm$ 0.003 & 6.575$^{+1.726}_{-1.547}$   & 0.32 &0.35 &0.33 &0.08 $\pm$ 0.01 &0.55$^{+0.15}_{-0.14}$\\
0.95 & \nodata & 0.066 $\pm$ 0.003 & 6.575$^{+1.726}_{-1.547}$   & 0.32 &0.35 &0.32 &0.12 $\pm$ 0.01 &0.81$^{+0.22}_{-0.19}$\\
1.05 & \nodata & 0.066 $\pm$ 0.003 & 7.390$^{+4.613}_{-5.688}$   & 0.32 &0.35 &0.32 &0.12 $\pm$ 0.01 &0.91$^{+0.57}_{-0.71}$\\
1.15 & \nodata & 0.102 $\pm$ 0.004 & 7.390$^{+4.613}_{-5.688}$  & 0.31 &0.35 &0.31 &0.20 $\pm$ 0.01 &1.46$^{+0.91}_{-1.12}$\\
\cutinhead{Stellar-mass selected pairs}
\hline
\sidehead{de Ravel et al. 2009; $R_{proj} < 100$ kpc $h^{-1}$; $1 < M_{primary}/M_{satellite} < 4$; $\delta V < 500$ km s$^{-1}$; log[$M_{star}$] $> 10 - 0.187z$}
0.52 & \nodata & 0.14 $\pm$ 0.06 & 4.11$^{+0.61}_{-0.54}$ & 1.61 & 1.52 & 1.54 & 0.052 $\pm$ 0.023 & 0.21 $\pm$ 0.10 \\
0.70 & \nodata & 0.26 $\pm$ 0.06 & 3.83$^{+0.56}_{-0.65}$ & 1.60 & 1.50 & 1.56 & 0.098 $\pm$ 0.022 & 0.37 $\pm$ 0.10\\
0.90 & \nodata & 0.25 $\pm$ 0.06 & 3.62$^{+0.55}_{-0.59}$ & 1.62 & 1.48 & 1.58 & 0.093 $\pm$ 0.023 & 0.34 $\pm$ 0.10\\
\hline
\sidehead{Bundy et al. 2009; $5 < R_{proj} < 20$ kpc $h^{-1}$; $1 < M_{primary}/M_{satellite} < 4$; $\delta z < 0.08(1 + z)$; $M_{star} > 10^{10} M_{\odot}$}
0.55 & \nodata  & 0.03 $\pm$ 0.02 & 4.22$\pm$ 0.70 & 0.33 & 0.35 & 0.34 & 0.055 $\pm$ 0.036 & 0.23 $\pm$ 0.16 \\
0.80 & \nodata  & 0.05 $\pm$ 0.03 & 4.71$\pm$ 0.60 & 0.33 & 0.35 & 0.33 & 0.091 $\pm$ 0.055 & 0.43 $\pm$ 0.26\\
1.15 & \nodata  & 0.06 $\pm$ 0.02 & 1.56$\pm$ 0.15 & 0.31 & 0.35 & 0.31 & 0.116 $\pm$ 0.039 & 0.18 $\pm$ 0.06\\
\enddata
\tablenotetext{a}{ $\Re_{merg}$ and $\Gamma_{merg}$ are calculated using Equations 3 and 6,  and $\langle T_{obs}(z) \rangle_{S08}$. We assume $C_{merge} = 0.6$
for all close pair samples.}
\end{deluxetable*}

\begin{deluxetable*}{lccccccc}
\tablecolumns{8}
\tabletypesize{\scriptsize}
\tablecaption{$G-M_{20}$ Merger Fractions and Merger Rates}
\tablehead{ \colhead{$z$} & \colhead{$f_{merg}$}   & \colhead{$n_{gal}$}  & \colhead{$\langle T_{obs} \rangle_{S08}$}  & \colhead{$\langle T_{obs} \rangle_{C06}$}  
& \colhead{$\langle T_{obs} \rangle_{St09}$}  & \colhead{$\Re_{merg}$\tablenotemark{a}}  & \colhead{$\Gamma_{merg}$\tablenotemark{a}} \\
 \colhead{}    &\colhead{}   &\colhead{[$10^{-3}$ Mpc$^{-3}$]} &\colhead{ [Gyr]}  &\colhead{ [Gyr]} &\colhead{ [Gyr]}  & \colhead{Gyr$^{-1}$} &\colhead{  [$10^{-3}$ Gyr$^{-1}$ Mpc$^{-3}$]}}
\startdata
\sidehead{Lotz et al. 2008; $M_B \leq -18.94 - 1.3z$}
0.3 &0.09 $\pm$ 0.03 &4.32$^{+0.30}_{-0.27}$ & 0.20 & 0.22 & 0.22 & 0.45 $\pm$ 0.15 & 1.94 $\pm$ 0.66 \\
0.5 &0.13 $\pm$ 0.04 &4.44$^{+0.14}_{-0.15}$ & 0.20 & 0.21 & 0.22 & 0.65 $\pm$ 0.20 & 2.89 $\pm$ 0.89\\
0.7 &0.07 $\pm$ 0.01 &4.00$^{+0.21}_{-0.18}$ & 0.20 & 0.21 & 0.22 & 0.35 $\pm$ 0.05 & 1.40 $\pm$ 0.21\\
0.9 &0.07 $\pm$ 0.01 &3.56$^{+0.10}_{-0.12}$ & 0.20 & 0.21 & 0.23 & 0.35 $\pm$ 0.05 & 1.25 $\pm$ 0.18\\
1.1 &0.13 $\pm$ 0.02 &2.49$^{+0.22}_{-0.26}$ & 0.19 & 0.21 & 0.23 & 0.68 $\pm$ 0.11 & 1.70$^{+0.30}_{-0.32}$\\
\hline
\sidehead{Lotz et al. 2008; $M_{star} \ge 10^{10} M_{\odot}$}
0.3 &0.07 $\pm$ 0.03 & 5.25 $\pm$ 0.24 & 0.20 & 0.21 & 0.22 & 0.37 $\pm$ 0.15 & 1.84 $\pm$ 0.79 \\
0.5 &0.13 $\pm$ 0.03 & 3.04 $\pm$ 0.12 & 0.20 & 0.21 & 0.22 & 0.64 $\pm$ 0.15 & 1.98 $\pm$ 0.46\\
0.7 &0.06 $\pm$ 0.02 & 3.16 $\pm$ 0.11 & 0.20 & 0.21 & 0.22 & 0.29 $\pm$ 0.10 & 0.95 $\pm$ 0.32\\
0.9 &0.08 $\pm$ 0.02 & 3.63 $\pm$ 0.09 & 0.20 & 0.21 & 0.23 & 0.39 $\pm$ 0.10 & 1.45 $\pm$ 0.36\\
1.1 &0.12 $\pm$ 0.02 & 2.33 $\pm$ 0.07 & 0.19 & 0.21 & 0.23 & 0.62 $\pm$ 0.11 & 1.47 $\pm$ 0.25\\
\hline
\sidehead{$M_B < -19.2$ ($n_{gal} \sim 6 \times 10^{-3} Mpc^{-3}$)}
0.3 &0.10 $\pm$ 0.02 & 4.77$^{+0.33}_{-0.29}$ & 0.20 & 0.21 & 0.22 & 0.52 $\pm$ 0.10 & 2.48$^{+0.51}_{-0.50}$ \\
0.5 &0.13 $\pm$ 0.02 & 6.01$^{+0.19}_{-0.20}$ & 0.20 & 0.21 & 0.22 & 0.66 $\pm$ 0.08 & 3.97$^{+0.50}_{-0.50}$\\
0.7 &0.08 $\pm$ 0.01 & 6.40$^{+0.34}_{-0.28}$ & 0.20 & 0.21 & 0.22 & 0.41 $\pm$ 0.04 & 2.59$^{+0.29}_{-0.28}$\\
0.9 &0.09 $\pm$ 0.01 & 7.34$^{+0.20}_{-0.25}$ & 0.20 & 0.21 & 0.23 & 0.47 $\pm$ 0.04 & 3.41$^{+0.31}_{-0.32}$\\
1.1 &0.12 $\pm$ 0.01 & 6.25$^{+0.55}_{-0.66}$ & 0.19 & 0.21 & 0.23 & 0.61 $\pm$ 0.05 & 3.78$^{+0.47}_{-0.52}$\\
\enddata
\tablenotetext{a}{ $\Re_{merg}$ and $\Gamma_{merg}$ are calculated using Equations 3 and 6,  and $\langle T_{obs}(z) \rangle_{S08}$.  }
\end{deluxetable*}

\begin{deluxetable*}{lccccccc}
\tablecolumns{8}
\tabletypesize{\scriptsize}
\tablecaption{Asymmetry Merger Fractions and Merger Rates}
\tablehead{ \colhead{$z$} & \colhead{$f_{merg}$}   & \colhead{$n_{gal}$} & \colhead{$\langle T_{obs} \rangle_{S08}$\tablenotemark{a}}  & \colhead{$\langle T_{obs} \rangle_{C06}$\tablenotemark{a}}  & \colhead{$\langle T_{obs} \rangle_{St09}$\tablenotemark{a}}  &\colhead{$\Re_{merg, S08}$\tablenotemark{b}} & \colhead{$\Gamma_{merg, S08}$\tablenotemark{b}} \\
 \colhead{}    &\colhead{}   &\colhead{[$10^{-3}$ Mpc$^{-3}$]}  &\colhead{ [Gyr]}  &\colhead{ [Gyr]} &\colhead{ [Gyr]} 
&\colhead{  [Gyr$^{-1}$]} &\colhead{  [$10^{-3}$ Gyr$^{-1}$ Mpc$^{-3}$]}}
\startdata
\sidehead{Conselice et al. 2009; $M_{star} \ge 10^{10} M_{\odot}$}
0.25 &0.04 $\pm$ 0.01 &5.25 $\pm$ 0.24 &0.39/0.46 &0.17/0.27 &0.24/0.30 & 0.10 $\pm$ 0.03 /0.09 $\pm$ 0.02 & 0.54 $\pm$ 0.14/ 0.46 $\pm$ 0.12\\
0.35 &0.04 $\pm$ 0.01 &5.25 $\pm$ 0.24 &0.41/0.47 &0.17/0.27 &0.28/0.33 & 0.10 $\pm$ 0.02 /0.09 $\pm$ 0.02 & 0.51 $\pm$ 0.13 /0.45 $\pm$ 0.11\\
0.45 &0.04 $\pm$ 0.01 &3.04 $\pm$ 0.12 &0.45/0.50 &0.18/0.28 &0.28/0.33 & 0.09 $\pm$ 0.02/ 0.08 $\pm$ 0.02 & 0.27 $\pm$ 0.07/ 0.24 $\pm$ 0.06\\
0.55 &0.04 $\pm$ 0.01 &3.04 $\pm$ 0.12 &0.47/0.52 &0.18/0.28 &0.31/0.35 & 0.09 $\pm$ 0.02/ 0.08 $\pm$ 0.02 & 0.26 $\pm$ 0.07 /0.23 $\pm$ 0.06\\
0.65 &0.09 $\pm$ 0.01 &3.16 $\pm$ 0.11 &0.53/0.57 &0.19/0.28 &0.34/0.38 & 0.17 $\pm$ 0.02/ 0.16 $\pm$ 0.02 & 0.54 $\pm$ 0.06 /0.50 $\pm$ 0.06\\
0.75 &0.12 $\pm$ 0.01 &3.16 $\pm$ 0.11 &0.53/0.56 &0.20/0.29 &0.34/0.38 & 0.23 $\pm$ 0.02/ 0.21 $\pm$ 0.02 & 0.72 $\pm$ 0.06 /0.68 $\pm$ 0.06\\
0.85 &0.11 $\pm$ 0.01 &3.63 $\pm$ 0.09 &0.62/0.64 &0.21/0.30 &0.35/0.38 & 0.18 $\pm$ 0.02 /0.17 $\pm$ 0.02 & 0.64 $\pm$ 0.06 /0.62 $\pm$ 0.06\\
0.95 &0.10 $\pm$ 0.01 &3.63 $\pm$ 0.09 &0.63/0.64 &0.21/0.30 &0.38/0.41 & 0.16 $\pm$ 0.02/ 0.16 $\pm$ 0.02 & 0.58 $\pm$ 0.06 /0.57 $\pm$ 0.06\\
1.05 &0.11 $\pm$ 0.01 &2.33 $\pm$ 0.07 &0.68/0.69 &0.22/0.31 &0.38/0.41 & 0.16 $\pm$ 0.01 /0.16 $\pm$ 0.01 & 0.38 $\pm$ 0.04/ 0.37 $\pm$ 0.04\\
1.15 &0.13 $\pm$ 0.01 &2.33 $\pm$ 0.07 &0.70/0.71 &0.23/0.31 &0.44/0.47 & 0.19 $\pm$ 0.01/ 0.18 $\pm$ 0.01 & 0.43 $\pm$ 0.04/ 0.43 $\pm$ 0.04\\
\hline
\sidehead{L\'{o}pez-Sanjuan et al. 2009; $M_{star} \ge 10^{10} M_{\odot}$}
0.4     & 0.006$\pm$ 0.02 & 4.22$\pm$ 0.70 & 0.43/0.49 &0.17/0.27 &0.27/0.33 &0.01 $\pm$ 0.05 /0.01 $\pm$ 0.04 &0.15$^{+0.49}_{-0.49}$/ 0.09$^{+0.31}_{-0.31}$\\
0.725 & 0.022$\pm$ 0.01 & 4.71$\pm$ 0.60 & 0.54/0.58 &0.19/0.29 &0.34/0.38 &0.04 $\pm$ 0.02/ 0.04 $\pm$ 0.02 &0.54$^{+0.25}_{-0.25}$/ 0.36$^{+0.17}_{-0.17}$\\
0.975 & 0.037$\pm$ 0.01 & 1.56$\pm$ 0.15 & 0.65/0.67 &0.21/0.30 &0.37/0.40 &0.06 $\pm$ 0.02/ 0.06 $\pm$ 0.01 &0.27$^{+0.08}_{-0.08}$/ 0.19$^{+0.05}_{-0.05}$\\
\hline
\sidehead{Shi et al. 2009; $M_B \le -18.94 - 1.3z$}
0.3 & 0.21 $\pm$ 0.04 &4.32$^{+0.30}_{-0.27}$ &0.40/0.47 &0.17/0.26 &0.26/0.32 &0.52 $\pm$ 0.10/ 0.45 $\pm$ 0.09 &2.27$^{+0.46}_{-0.45}$ /1.93$^{+0.39}_{-0.39}$ \\
0.5 & 0.24 $\pm$ 0.03 &4.44$^{+0.14}_{-0.15}$ &0.46/0.51 &0.18/0.27 &0.29/0.34 &0.52 $\pm$ 0.07/ 0.47 $\pm$ 0.06 &2.32$^{+0.30}_{-0.30}$ /2.09$^{+0.27}_{-0.27}$\\
0.7 & 0.25 $\pm$ 0.02 &4.00$^{+0.21}_{-0.18}$ &0.53/0.56 &0.19/0.29 &0.34/0.38 &0.47 $\pm$ 0.04/ 0.45 $\pm$ 0.04 &1.89$^{+0.18}_{-0.17}$ /1.79$^{+0.17}_{-0.16}$\\
0.9 & 0.36 $\pm$ 0.03 &3.56$^{+0.10}_{-0.12}$ &0.62/0.64 &0.21/0.30 &0.36/0.40 &0.58 $\pm$ 0.05/ 0.56 $\pm$ 0.05 &2.07$^{+0.18}_{-0.19}$ /2.00$^{+0.18}_{-0.18}$\\
1.1 & 0.30 $\pm$ 0.03 &2.49$^{+0.22}_{-0.26}$ &0.69/0.70 &0.23/0.32 &0.41/0.44 &0.43 $\pm$ 0.04/ 0.43 $\pm$ 0.04 &1.08$^{+0.14}_{-0.16}$ /1.07$^{+0.14}_{-0.15}$\\
\enddata
\tablenotetext{a}{ $\langle T_{obs}\rangle$ calculated assuming both $T_{obs}(A)$ = 0 for $f_{gas} < 0.1$,  and $T_{obs}(A)$ at $f_{gas} < 0.1$ is the same 
as $T_{obs}(A)$ at $f_{gas} \sim 0.2$. } 
\tablenotetext{b}{ $\Re_{merg}$ and $\Gamma_{merg}$ are calculated using Equations 3 and 6,  and $\langle T_{obs}(z) \rangle_{S08}$. }
\end{deluxetable*}

\subsection{The Major Merger Rate: Close Pairs}
The close pair studies included in this paper select
galaxies with luminosity or stellar mass ratios less than
1:2 or 1:4. Although luminosity brightening of an interacting 
satellite galaxy may cause the measured luminosity ratio to 
be less than the stellar or baryonic mass ratio
(e.g. Bundy et al. 2004), we assume that this sample primarily 
probes ‘major mergers’ with baryonic mass ratios
roughly comparable to their luminosity or stellar mass
ratios.

When samples with similar parent selection criteria are
compared, we find that the merger rates derived from
various close pair studies are remarkably consistent (Figure 10). 
However, the evolution of the ‘merger rate’
depends on whether one calculates the merger rate per
galaxy ($\Re$) or the merger rate per unit volume ($\Gamma$). The
best-fit evolution in $\Gamma_{pairs}(z)$ is weaker than $\Re_{pairs}(z)$
because the evolution in $n_{gal}(z)$ is opposite to the trend
in $\Re_{pairs}(z)$ (see Figures 1 and 2). The major merger rate
($\Gamma_{pairs}$ or $\Re_{pairs}$) and its evolution with redshift are 
similar for the stellar-mass and evolving luminosity selected
samples, suggesting that luminosity-brightening does not
significantly bias the luminosity-selected merger samples.
In the left hand side of Figure 10, we plot $\Gamma$ and $\Re$ for
1:1 - 1:4 pairs with $M_{star} > 10^{10} M_{\odot}$ from the Bundy et
al. (2009) study (blue circles) and de Ravel et al. (2009)
study (black circles). Despite being drawn from different
fields with different close pair criteria, these agree well
once the corresponding observability timescales are applied. The best
fit volume-averaged merger rate $\Gamma_{pairs, M_{star}}(z)$ 
(blue line, top left panel)  and  fractional merger rate
$\Re_{pairs,M_{star}}(z)$ (blue line, bottom left panel) are given in Table 4.

On the right hand side of Figure 10, we plot $\Gamma$ and
$\Re$ for close pairs selected with evolving luminosity cuts
from Lin et al. 2008 (cyan circles), de Ravel et al.
2009 (black circles), and Kartaltepe et al. 2007 (blue
circles), and the $z \sim 0.1$ value from the Patton \& Atfield 
2008 study (red circle). Most of these studies give very 
consistent galaxy merger rates once corrected for the 
observability timescales, although the de Ravel et al (2009) 
luminosity-selected bright pairs have a merger rate $\sim$ 3 
times higher than the other studies at $z \sim 0.7 - 1$. 
However, the number density of galaxies with the passive luminosity
evolution (PLE) assumptions adopted by de
Ravel et al. (2009) is also $\sim 2-3$ times higher than
the selection adopted by the other studies. As we discuss in 
\S5.4, the de Ravel et al. (2009) luminosity-selected pairs are 
in good agreement with other studies selected at similar number 
densities. We exclude the de Ravel et al. (2009) pairs and give 
the best fit volume-averaged merger rate for the luminosity-selected close pairs
$\Gamma_{pairs,PLE}(z)$ (blue line, top right panel)  and  fractional merger rate  
$\Re_{pairs,PLE}(z)$ (blue line, bottom right panel) in Table 4.

\begin{figure*}
\plotone{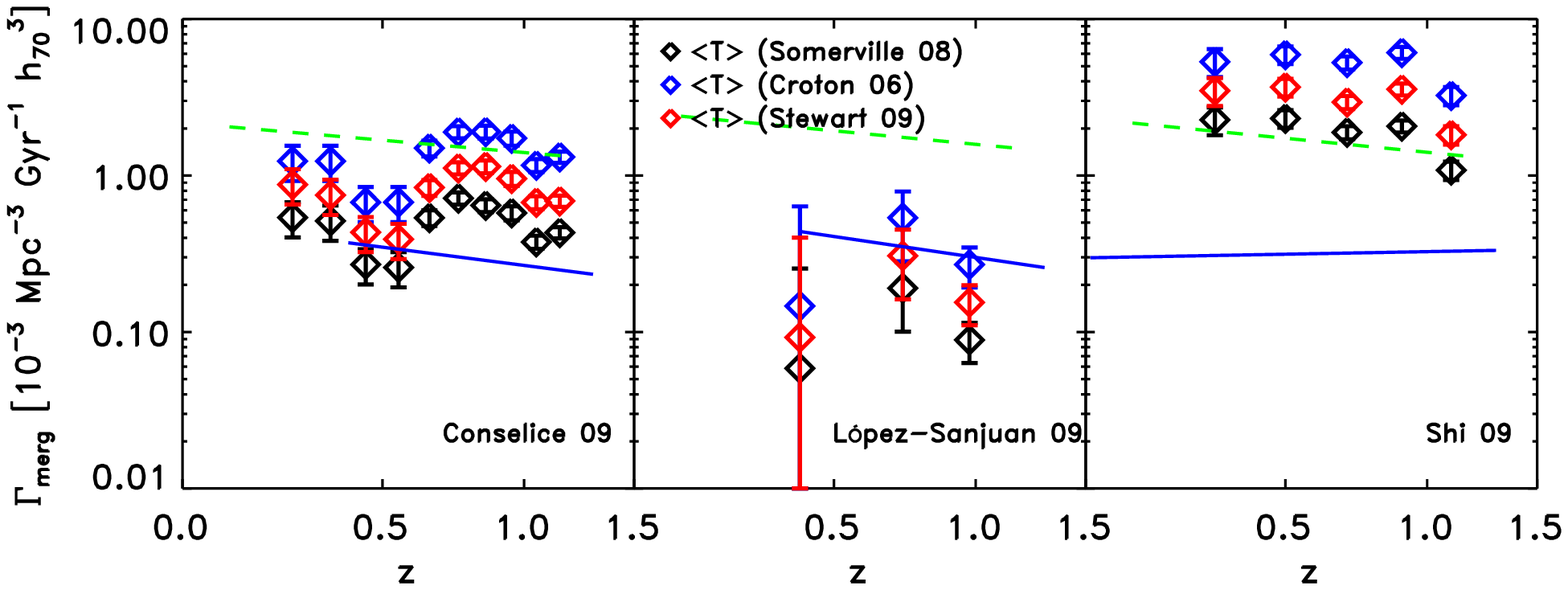}
\plotone{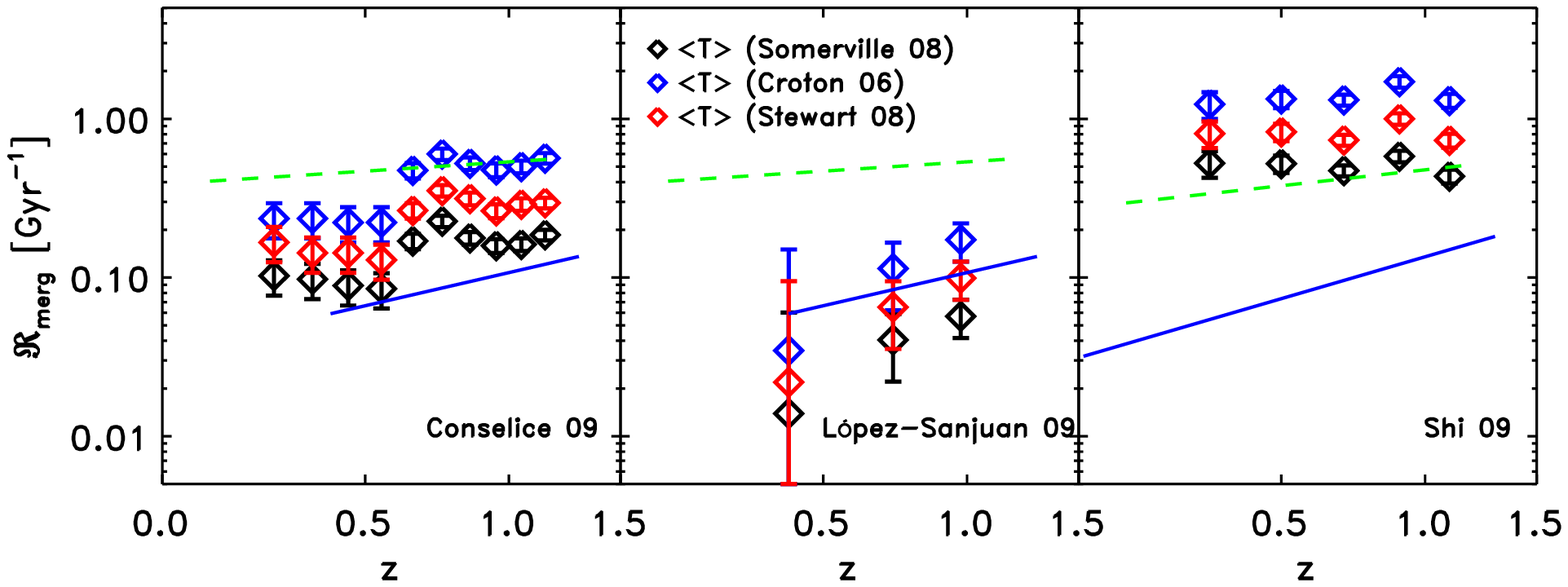}
\caption{ Top: $\Gamma_{merg}$, the merger rate per co-moving unit volume for asymmetric galaxies, 
for Conselice et al. 2009 and L\'{o}pez-Sanjuan et al. 2009 stellar mass-selected samples (left, center) and 
the Shi et al. 2009 rest-frame $B$ luminosity selected sample
(right). Bottom: $\Re_{merg}$, the fractional merger rate for the same samples. $\Gamma_{merg}$ and $\Re_{merg}$ 
are calculated using the three different observability timescales (Somerville et al. 2008, black; Croton et al. 2006, 
blue; Stewart et al. 2009b, red) for each study. For comparison
are the best-fit to the major merger rates from the close pairs (blue solid lines) and major+minor merger rates 
from $G-M_{20}$ (green dashed lines) from Figure 10. The error-bars are computed using the observational uncertainties on $f_{merg}$  and $n_{gal}$ and
do not include uncertainties in $\langle T_{obs} \rangle$.}
\end{figure*}

\subsection{ $G-M_{20}$ and the Minor Merger Rate }
We also calculate $\Gamma$ and $\Re$ from the $G - M_{20}$ 
merger fractions for the stellar-mass and evolving luminosity selected 
parent samples (Table 2; green points in Figure 10). Both $\Gamma_{G-M_{20}}$ 
and $\Re_{G-M_{20}}$ are significantly higher
than the merger rates calculated for close pairs. The
best-fit evolution with redshift in the $G-M_{20}$ merger
rates is systematically weaker than that for the close
pairs, but have slopes that are consistent within the large
uncertainties. As we discussed in \S4, $G-M_{20}$ is sensitive 
to mergers with baryonic mass ratios between 1:1 and 1:10, while 
the close pair studies plotted in Figure 10 have mass ratios ∼ 1:1 - 1:4. 
Therefore the $G-M_{20}$ ‘merger rate’ includes minor as well as major mergers,
and is expected to be significantly higher than the major merger rate.

We give the best-fit $G-M_{20}$-derived volume-averaged `total' merger rate 
$\Gamma_{G-M_{20}}$  and `total' merger rate per galaxy  $\Re_{G-M_{20}}$ 
for stellar mass selected and luminosity-selected parent samples in 
Table 4 and Figure 10 (green dashed lines).  
The $G-M_{20}$ merger rates evolve weakly with redshift,
and show trends consistent with the pair merger rate evolution
within the large uncertainties.  However, 
the $G-M_{20}$ merger rates are much higher than
the corresponding close pair merger rates at $z  \sim  0.7$,
where the merger rates are best determined. We can
estimate the minor merger rate for galaxies with mass
ratio between 1:4 and 1:10 by subtracting the best-fit 
close-pair derived major merger rate from the total
$G-M_{20}$ merger rate.  These minor merger rates are
given in Table 4. 

We find that the fractional and volume-averaged minor
merger rate at $z \sim 0.7$ is $\sim 3$ times that of the major
merger rate for galaxies selected by stellar mass or PLE.
This ratio does not evolve significantly with redshift between $0.2 < z < 1.2$. 
Our findings are roughly consistent with the relative numbers of minor/major mergers from the
visual classification study of Jogee et al. (2009), who find
$\sim$ 3 minor mergers for every major merger for galaxies
selected at $0.24 < z < 0.8$ and $M_{star} > 2.5 \times 10^{10} M_{\odot}$.
However, the absolute value of the major+minor merger
rate per co-moving volume ($\Gamma$) calculated in Jogee et al.
(2009) is a factor of ten lower than what we find
here, in part because of the higher mass limit of their
sample.  Weaker evolution of minor mergers relative to major
mergers is also seen in the close pair study at $z < 1$ by L\'{o}pez-Sanjuan et al. (2011).
In this case, however,  L\'{o}pez-Sanjuan et al. (2011) find 
only $0.5 - 1$ minor merger pairs (with rest-frame $B$-band luminosity ratios between 1:10 and
1:4) per major merger pair.    Finally, we note that our estimates of the
minor merger rate are contingent upon the accuracy of both our close-pair determined
major merger rate and $G-M_{20}$ determined 'total' merger rates. 

\subsection{Asymmetry Merger Rates}
Both observational and theoretical uncertainties make
it difficult to  calculate the merger rate reliably with
asymmetric galaxies. The asymmetric fraction measured
by three different groups for the same data vary by a
factor of ten, reflecting different corrections for surface-brightness 
dimming and contamination by non-mergers.
Furthermore, the timescales needed to compute merger rates
also vary by factors of 2-3, depending on the assumed
distribution and evolution of gas fractions. The high-resolution 
merger simulations predict that asymmetry is
sensitive to mergers with mass ratios less than 1:4 for local 
gas fractions, and 1:10 for high gas fractions. Therefore, 
in order to be consistent with the results from close
pairs and $G-M_{20}$ mergers, the merger rate from asymmetry 
should lie between the close pair derived major merger rate and 
the $G-M_{20}$ major + minor merger rate.

We calculate $\Gamma_{A}(z)$ and $\Re_A(z)$ (Table 3, Figure
11), using the asymmetric galaxy fraction measurements
from Conselice et al. 2009 (left panel), L\'{o}pez-Sanjuan et
al. 2009 (center panel), and Shi et al. 2009 (left panel)
and $\langle T_{obs}(A, z) \rangle$ predicted by the S08 (black points),
C06 (blue points), and St09 (red points) models from Figure
9. The uncertainty in $\langle T_{obs}(A) \rangle$ results in a factor
of 2-3 uncertainty in $\Gamma$ and $\Re$ for a given set of asymmetry measurements.
For a given $\langle T_{obs}(A) \rangle$, the calculated 
$\Gamma_A$ and $\Re_A$ vary by a factor of ten for
the different asymmetry studies. The volume-averaged
asymmetry-derived merger rates $\Gamma_A$ have weak evolution with redshift, 
with best-fit $\alpha$ values ranging from $-1$
to $+0.4$. The fractional asymmetry-derived merger rates
$\Re_A$ evolve more strongly with redshift, with best-fit
$m$ values ranging from $+0.1$ to $+1.6$. (We exclude the
fits to the L\'{o}pez-Sanjuan et al. points because of their
large uncertainties.)

We compare the asymmetry-derived merger rates to
the close pair major merger (blue solid lines) and $G-M_{20}$
major+minor merger rates (green dashed lines) computed 
for parent samples with similar selections. If the
distribution of merger properties predicted by the S08
models is correct, then asymmetry should select many
gas-rich minor mergers and be expected to give $\Gamma_A(z)$
and $\Re_A(z)$ similar to those calculated with $G-M_{20}$
(green dashed lines). We find this to be the case for the
Shi et al. observations (black diamonds, right panels),
while the Conselice et al. observations and L\'{o}pez-Sanjuan 
et al. observations with S08 timescales give merger
rates significantly lower than the $G-M_{20}$ merger rate
(black diamonds, left and center panels). On the other
hand, if most mergers have low gas fractions at $z \sim 1$
(C06), then $\Gamma_A(z)$ and $\Re_A(z)$ are expected to
probe 1:1 - 1:4 mass ratio mergers and be consistent with
the close pair major merger rate (blue solid lines). The
L\'{o}pez Sanjuan et al. observations + C06 timescales are
roughly consistent with the close-pair major merger rates
(blue diamonds, center panels). Finally, if the intermediate 
gas-evolution scenario of St09 is correct, then asymmetric 
galaxies are more likely to be mergers
with intermediate mass ratios between 1:4 and 1:10, as
suggested by the Conselice et al. observations + St09
timescales (red diamonds, left panels). Until better observational
 constraints on the distribution and evolution in $f_{gas}$ are available, 
we cannot determine if (or which)
asymmetry merger rate calculations are in agreement
with other methods for measuring the galaxy merger
rate.

\begin{figure*}
\epsscale{1.2}
\plotone{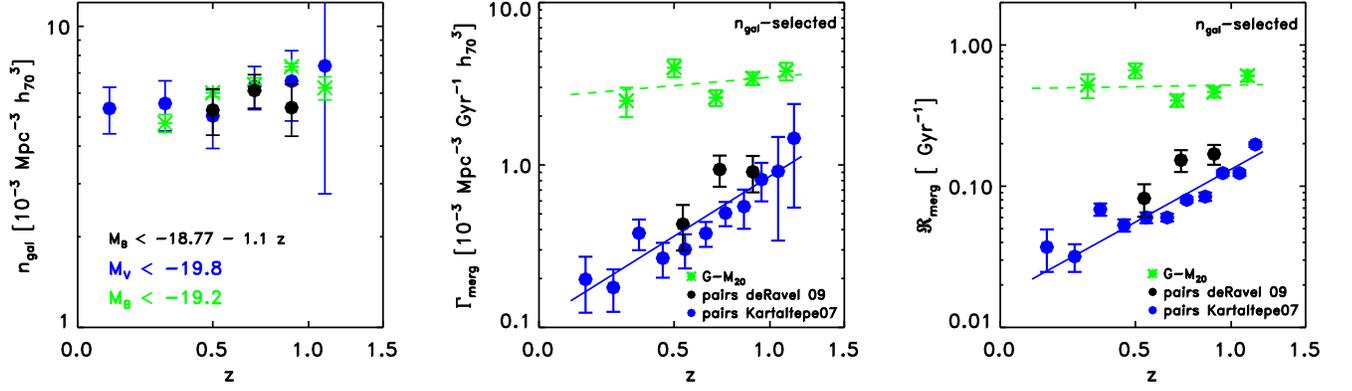}
\caption{ Left: The Kartaltepe et al. (2007; blue points) and de Ravel et al. (2009; black points) parent galaxy selections result in a
sample of galaxies with roughly constant $n_{gal}$ from $0.1 < z < 1.2$. For comparison, we select $G-M_{20}$ merger candidates with a
fixed $M_B$ limit $\le -19.2$ (green asterisks) in order to match the roughly constant $n_{gal}$ of these studies. 
Center: The galaxy merger rates per unit co-moving volume for samples selected at constant $n_{gal}$ v. redshift. 
. Right:  The galaxy merger rates per galaxy for samples selected
at constant $n_{gal}$ v. redshift.   We find strong evolution in $\Gamma_{pairs, n_{gal}}$ and $\Re_{pairs, n_{gal}}  \propto (1 + z)^{3}$, 
and weaker evolution in the `total' and inferred minor merger rates (see Table 4). 
The error-bars are computed using the observational uncertainties on $f_{merg}$, $f_{pair}$,  and $n_{gal}$ and
do not include uncertainties in $\langle T_{obs} \rangle$. }
\epsscale{1.}
\end{figure*}

\begin{figure*}
\plotone{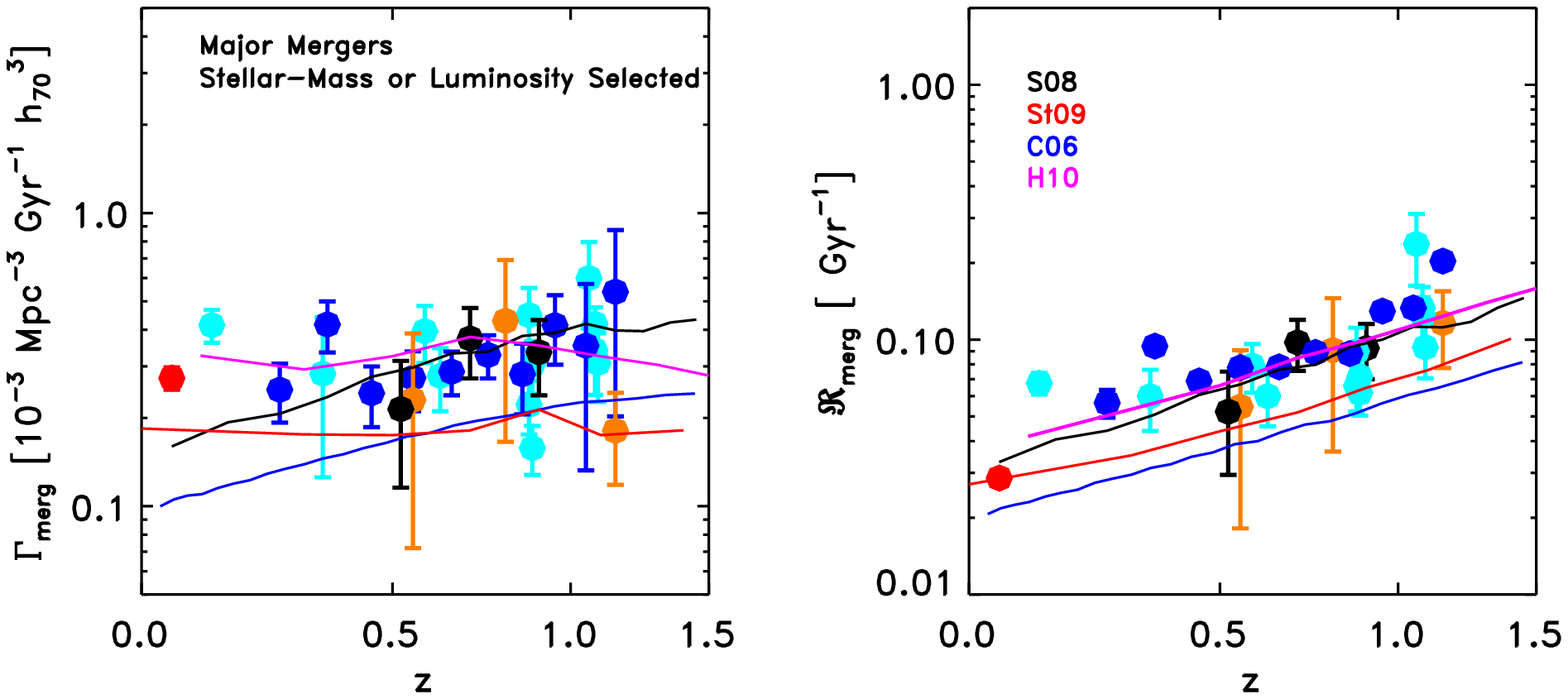}
\plotone{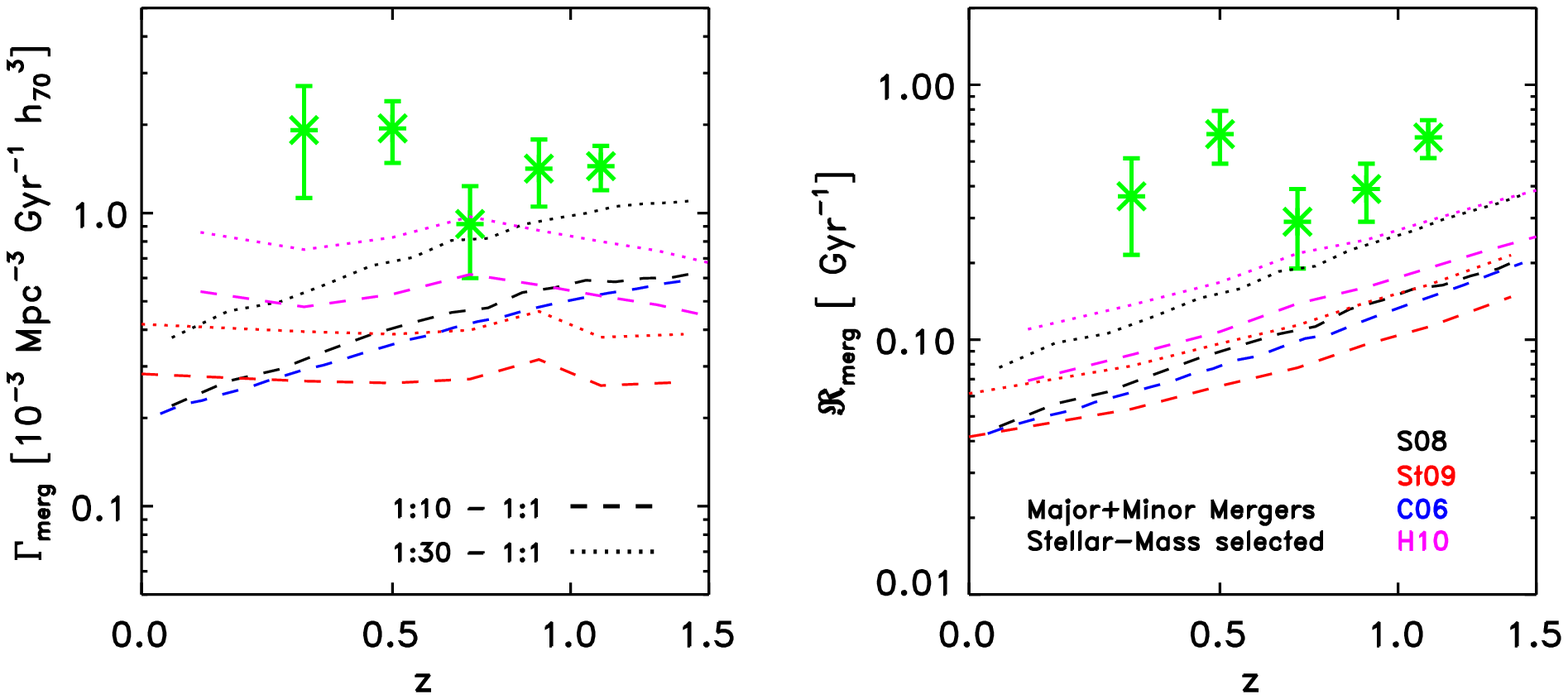}
\caption{ Observed Galaxy Merger Rates v. Theoretical Predictions. {\it Top:}  The volume-averaged (top left)
and fractional major merger (top right) rates given by stellar-mass and luminosity-selected 
close pairs are compared to the major merger rates given by the S08 (black lines), St09 (red lines), C06
(blue line),  and Hopkins et al. 2010b (magenta lines) models for 1:1 - 1:4 stellar mass ratio mergers and galaxies with $M_{star} > 10^{10}$ $M{_\odot}$.  The theoretical predictions are in good agreement with the observed major merger rates.  {\it Bottom:} The volume-averaged (bottom left)
and fractional major + minor merger (bottom right) rates given by stellar-mass selected 
$G-M_{20}$ mergers are compared to the major + minor merger rates given by  the same models for 1:1 - 1:10 (dashed lines) and 1:1 - 1:30 (dotted lines) stellar mass ratios and galaxies with $M_{star} > 10^{10}$ $M{_\odot}$.   The observed $G-M_{20}$ `total' merger rates are an order of magnitude higher than the 
predicted 1:1-1:10 merger rates,  and a factor of ~2-3 times higher than the predicted 1:1-1:30 merger rates.  }
\end{figure*}

\begin{deluxetable}{lll}
\setlength{\tabcolsep}{0.01in}
\tablewidth{0pc}
\tablecolumns{3}
\tabletypesize{\scriptsize}
\tablecaption{Major and Minor Galaxy Merger Rates at $z < 1.5$}
\tablehead{ \colhead{Selection} & \colhead{Mass Ratio}   & \colhead{ $C \times (1+z)^{\alpha}$ }     }
\startdata
\cutinhead{ Volume-Averaged Total Merger Rates [$10^{-3}$ Gyr$^{-1}$ Mpc$^{-3}$  h$_{70}^3$]}
$\Gamma_{G-M_{20}, M_{star}}(z)$  & 1:1 - 1:10 (30) & $(2.2 \pm 2.7)    (1+z)^{-0.6 \pm 1.6}$ \\
$\Gamma_{G-M_{20}, PLE}(z)$       & ...                               & $(2.3 \pm 2.5)    (1+z)^{-0.7 \pm 1.5}$ \\
$\Gamma_{G-M_{20}, n_{gal}}(z)$     & ...                         & $(2.60 \pm 0.06)    (1+z)^{+0.4 \pm 0.9}$ \\
\cutinhead{ Total Merger Rates per galaxy [ Gyr$^{-1}$]}
$\Re_{G-M_{20}, M_{star}}(z)$   & 1:1 - 1:10 (30)          & $(0.4 \pm 0.4)    (1+z)^{+0.5 \pm 1.6}$ \\
$\Re_{G-M_{20}, PLE}(z)$        &  ...                                 & $(0.3 \pm 0.2)    (1+z)^{+0.8 \pm 1.4}$ \\
$\Re_{G-M_{20}, n_{gal}}(z)$    &  ...                                & $(0.5 \pm 0.3)    (1+z)^{+0.1 \pm 0.8}$ \\
\cutinhead{ Volume-Averaged Major Merger Rates [$10^{-3}$ Gyr$^{-1}$ Mpc$^{-3}$  h$_{70}^3$]}
$\Gamma_{pairs, M_{star}}(z)$ & 1:1 - 1:4  & $(0.5 \pm 1.4)    (1+z)^{-0.9 \pm 3.2}$ \\
$\Gamma_{pairs, PLE}(z)$      & ...  & $(0.30 \pm 0.04)    (1+z)^{+0.1  \pm 0.4}$ \\
$\Gamma_{pairs, n_{gal}}(z)$     & ...  & $(0.11 \pm 0.03)    (1+z)^{+3.0 \pm 1.1}$ \\
\cutinhead{ Major Merger Rates per galaxy [ Gyr$^{-1}$]}
$\Re_{pairs, M_{star}}(z)$   & 1:1 - 1:4 & $(0.03 \pm 0.01)    (1+z)^{+1.7 \pm 1.3}$ \\
$\Re_{pairs, PLE}(z)$        &  ...                   & $(0.03 \pm 0.01)    (1+z)^{+2.11  \pm 0.2}$ \\
$\Re_{pairs, n_{gal}}(z)$  &  ...             & $(0.016 \pm 0.001)    (1+z)^{+3.0 \pm 0.3}$ \\
\cutinhead{ Volume-Averaged Minor Merger Rates [$10^{-3}$ Gyr$^{-1}$ Mpc$^{-3}$  h$_{70}^3$]}
$\Gamma_{minor, M_{star}}(z)$   & 1:4 - 1:10 (30)  & $(0.8 \pm 5.1)    (1+z)^{-0.2 \pm 2.8}$ \\
$\Gamma_{minor, PLE}(z)$         & ...                               & $(1.0 \pm 0.1)    (1+z)^{-0.3 \pm 1.0}$ \\
$\Gamma_{minor, n_{gal}}(z)$     & ...                       & $(2.6 \pm 1.1)    (1+z)^{+0.1 \pm 1.6}$ \\
\cutinhead{ Minor Merger Rates per galaxy [ Gyr$^{-1}$]}
$\Re_{minor, M_{star}}(z)$   & 1:4 - 1:10 (30)           & $(0.27 \pm 0.08)    (1+z)^{-0.1 \pm 0.7}$ \\
$\Re_{minor, PLE}(z)$        &  ...                                       & $(0.28 \pm 0.08)    (1+z)^{-0.1 \pm 0.6}$ \\
$\Re_{minor, n_{gal}}(z)$    &  ...                                & $(0.37 \pm 0.03)    (1+z)^{-0.2 \pm 0.2}$ 
\enddata
\tablenotetext{}{The lower limit of the baryonic mass ratio for $G-M_{20}$ detected mergers is between 1:10 and
1:30.  See \S 5 for discussion. }
\end{deluxetable}

\subsection{Merger Rates for galaxies at constant $n_{gal}(z)$}
As we discussed in \S3.5, the derived evolution in the
galaxy merger rate depends in large part on which galaxies
are included in the sample. It is clear from Figure 2 that
selecting galaxy samples above a fixed stellar mass or with a
PLE assumption does not track the same progenitor-descendant
populations. A better (although still not perfect) way to
track the same population of galaxies (and dark matter
halos) with redshift is to select galaxies above a fixed
number density. No galaxy merger studies published to
date have been designed to use a constant number density selection. 
However, we find that the fixed rest-frame $V$
luminosity cut employed by Kartaltepe et al.  (2007) close
pair study results in a roughly constant number density 
selection (Figure 12, left). Kartaltepe et al. (2007)
selected their parent sample above
a fixed absolute luminosity for $0.1 < z < 1.2$, arguing that 
galaxy pairs and mergers would not follow a simple passive 
luminosity evolution because of merger-induced star-formation. 
The luminosity cut adopted by de
Ravel et al. ($M_B < -18.77 + 1.1z$) also selects 
a roughly constant and similar number density of
galaxies (black points). In order to compare the $G-M_{20}$
results to these studies, we adopt a fixed rest-frame $B$ 
luminosity cut of $M_B$(AB) $< -19.2$ which selects galaxies
at a similar number density at each redshift bin (Figure
12, left). We compute the $G-M_{20}$ fraction and merger
rates for galaxies selected at $M_B$(AB) $< -19.2$ (Table 2,
Figure 12).

When a roughly constant co-moving number density ($n_{gal} \sim 6 \times 10^{-3}$
Mpc$^{-3}$) selection is applied to the parent sample, the
close-pair derived galaxy merger rates show strong evolution 
with redshift, as concluded by Kartaltepe et al.
(2007). Combining the Kartaltepe et al. (2007) and de
Ravel et al. (2009) samples (Table 1), we find the galaxy
merger rate per unit co-moving volume $\Gamma_{pairs,n_{gal}}$
(blue line, center panel) evolves much more strongly  ($\alpha = +3.0 \pm 1.1$) 
than than the merger rate derived for the PLE parent sample with
a declining number density ($\alpha = +0.1 \pm 0.4$;  see Table 4). 
The evolution in fractional merger rate $\Re_{pairs,n_{gal}}$ is similar to the evolution in 
volume-averaged merger rate. The best-fit to the $G-M_{20}$ major+minor rates for
galaxies selected by number density have systematically
weaker evolution (although the uncertainties are large). 
At $z \sim 0.7$, the volume-averaged and fractional minor
merger rates for galaxies selected at $n_{gal} \sim 6 \times 10^{-3}$ Mpc$^{-3}$ 
are $\sim 4-5$ times the major merger rates (Table 4).

\subsection{Theoretical Predictions for Major and Minor Galaxy Merger Rates}

We compare the theoretical predictions for the galaxy merger rates to the major and 
major+minor merger rates for the stellar-mass selected samples in Figure 13.  
Because we have used only the {\it relative} distribution of merger properties from the
S08, St09, and C06 models to calculated $\langle T_{obs}(z) \rangle$,  
the predictions of the volume-averaged and fractional merger
rates from those models are independent of our timescale calculations.  We do incorporate
the relative numbers of minor and major mergers into our timescale calculations,  but in
practice only the asymmetry timescales are strongly dependent on the assumed mass ratio
distributions.   We also compare
to the Hopkins et al. (2010b) semi-empirical model predictions using their 
{\rm merger$\_$rate$\_$calculator.pro} IDL routine.   We calculate the predicted merger rates for galaxies at $0.1 < z < 1.5$ with 
stellar masses $M_{star} \ge 10^{10} M_{\odot}$ and stellar mass ratios between 1:1-1:4 (solid lines, upper
panels in Fig. 13), 1:1-1:10 (dashed lines, lower panels),  and 1:1-1:30 (dotted lines, lower panels).  

We find that predicted major merger rates agree within a factor of $\sim 2$ of each other, and are in
good agreement with the close pair-derived major merger rates (top panels, Fig. 13).   
(For completeness, we have also plotted the luminosity-selected close pairs as these give 
the same merger rate as the stellar-mass selected pairs).    The evolution in the volume-averaged 
major merger rate $\Gamma_{merg}$ 
is somewhat weaker and in better agreement with the data for the semi-empirical models (St09, H10).  
 This is likely because those models use $n_{gal}(z)$ as an input parameter rather than give independent
predictions for $n_{gal}(z)$ as the S08 and C06 models do.   
 The evolution in the fractional major merger rate $\Re_{merg}$  is similar for all the models,  and in 
excellent agreement with the data. 

On the other hand,  the predicted total merger rates (major+minor) are significantly lower than 
what is observed for $G-M_{20}$ mergers.  Given the uncertainty in the lower mass ratio limit
for $G-M_{20}$-selected mergers,  we show both the 1:1-1:10 and 1:1-1:30 total merger rates.  
The 1:1-1:10 merger rates are a factor of ~5 less than
the $G-M_{20}$-derived merger rates,  while the 1:1-1:30 merger rates are a factor $\sim$ 2-3 lower. 
The evolutionary trends with redshift are similar to the major merger predictions,  with weaker 
evolution in $\Gamma_{merg}$ for the semi-empirical models that follows the same evolutionary 
trend as the data.   

There are several possible reasons for the discrepancy in normalization of the
minor merger rates.   The $G-M_{20}$ merger rates could be
over-estimated,  either because it suffers from large contamination of non-merging systems 
or because  we have  under-estimated $\langle T_{obs}(z) \rangle$.    Alternatively,  the models could
under-estimate the frequency of minor mergers.   A primary galaxy of stellar mass $\sim 10^{10} M_{\odot}$ 
undergoing a 1:10 merger will have a merging satellite galaxy with a stellar mass of $\sim 10^9 M_{\odot}$;  a 
1:30 merger will have  a merging satellite will have a stellar mass of  $3 \times 10^{8} M_{\odot}$.  
The total halo masses are roughly 10 times higher,  but are still approaching the numerical resolution 
of merger trees based on numerical simulations with typical particle masses 
$\sim 3 \times 10^8 M_{\odot}$ (e.g. St09, C06, H10).   Also, low mass satellite galaxies are 
strongly effected by difficult-to-model physics,  
including supernova feedback and satellite disruption.  Therefore it is possible that the 
simulations have incompletely sampled the minor merger populations (e.g. Hopkins et al. 2010a).     
Both higher resolution
models and improved estimates of the minor merger rates are needed to resolve this discrepancy. 

\section{SUMMARY}
We attempt to reconcile the disparate observational estimates of
the galaxy merger rate at $z < 1.5$,  and calculate the
fractional (per galaxy)  and volume-averaged (per co-moving volume)
galaxy merger rates for major (1:1 - 1:4)  and minor (1:4 - 1:10) 
mergers.   When physically-motivated cosmologically-averaged timescales, similar
parent sample selections, and consistent definitions of the
merger rate are adopted, we are able to derive self-consistent estimates of the major 
and minor galaxy merger rate and its evolution.  We conclude that the differences in the observed galaxy
merger fractions may accounted for by the different observability
timescales,  different mass-ratio sensitivities, and different parent galaxy
selections.

We compute the first cosmologically-averaged observability timescales for 
three different approaches to identifying mergers -- close pairs, $G-M_{20}$ and asymmetry -- using the
results of a large suite of high-resolution N-body/hydrodynamical merger simulations 
and three different  cosmological galaxy evolution models to predict the distribution of
galaxy merger mass ratios and baryonic gas fractions.
The timescales for close pairs and $G-M_{20}$ are relatively 
insensitive to the assumptions about the distribution of galaxy 
merger properties. However, the cosmologically-averaged asymmetry timescales vary by a 
factor of $2-3$ with different assumptions for the distribution and evolution of 
baryonic gas fractions.  In particular, if the mean gas fraction of merging galaxies
evolves as strongly as recent observations suggest, then
the typical timescale for identifying a galaxy merger via
asymmetry may increase by a factor of 2 from $z \sim 0$ to
$z \sim 1.5$.  We apply these timescales to the close pair fraction and
morphologically-determined merger fractions at $z < 1.5$
published in the recent literature to compute merger rates.   We
estimate the minor merger rate by computing the difference between the
close pair-derived and $G-M_{20}$-derived merger rates. 

We find the following: 

(1)  The evolution of the major and minor merger rates of
galaxies depends upon the definition of merger rate (fractional or 
volume-averaged) and the selection of the parent sample of galaxies 
for which the merger rate is measured.  For parent samples of fixed 
stellar mass or with an assumption of passive luminosity evolution (PLE),
the evolution in the volume-averaged merger rate $\Gamma_{merg}$
[Gyr$^{-1}$ Mpc$^{-3}$] is significant weaker than the evolution
in the fractional merger rate $\Re_{merg}$ [Gyr$^{-1}$]. This
is because the co-moving number density of galaxies with these selections 
declines by a factor of $2-3$ from $z \sim 0$ to $z \sim 1.5$.

(2) The fractional and volume-averaged major merger
rates at $0 < z < 1.5$ for galaxies selected above a fixed stellar 
mass or assuming passive luminosity evolution are calculated
for close pairs with stellar mass ratios or optical luminosity 
ratios between 1:1 and 1:4 (Tables 1 and 4).
The number of major mergers per bright/massive galaxies 
increases as $\sim (1+z)^2$ with redshift, but the number
of major mergers per co-moving volume (selected above a
constant stellar mass) does not evolve significantly with
redshift. The major merger rates for galaxies selected
with $M_{star} \ge 10^{10} M_{\odot}$ and with $L_B > 0.4 L^*_B$
agree within the uncertainties, implying that luminosity brightening 
does not strongly bias the (evolving) luminosity-selected sample.

(3) The $G-M_{20}$ derived merger rates for similarly-
selected parent samples are significantly larger than the
close-pair derived major merger rates (Tables 2 and 4). This is consistent with 
the simulation predictions that $G-M_{20}$ selected mergers 
span a wider range of mass ratio (1:1- 1:10) than the close pair 
major-merger studies (1:1 -1:4). We estimate the minor merger 
(1:4 $ > M_{satellite}:M_{primary} > $ 1:10) rate by subtracting the best-fit 
close pair merger rate from the $G-M_{20}$ merger rate (Table 4). 
The fractional and volume-averaged minor
merger rates are $\sim 3$ times the major merger rates at
$z \sim 0.7$, and suggest weaker evolution in the minor
merger rate than the major merger rate at $z < 1$.

(4) The merger rates derived from observations of the
fraction of asymmetric galaxies are highly uncertain (Table 3).
Different methodologies for correcting asymmetry for
surface-brightness dimming, sky noise, and  false
merger contamination result in a factor of 10 discrepancy 
in the literature values of the fraction of asymmetric
galaxies. The theoretical average observability timescales add a
factor of a few uncertainty to the asymmetry merger rate, 
because of their dependence on the unknown distribution of $f_{gas}$. 
The asymmetry-derived merger rates generally fall between the close-pair major 
merger rates and the $G-M_{20}$ major+minor rates, as
expected for the range of predicted merger mass ratios for
asymmetric galaxies ($> 1:4 - 1:10$).

(5) We have also measured the galaxy merger rates for samples selected
with roughly constant co-moving number density ($n_{gal}(z) \sim 6 \times 10^{-3}$ Mpc$^{-3}$). 
This selection results in stronger evolution in the merger rate,  because it probes
fainter/less massive galaxies at higher $z$ and brighter/more massive galaxies at low $z$
than fixed stellar mass selection.  This selection also is expected to better 
track progenitor-descendant populations, and better match dark matter halo merger rate
calculations.   Strong evolution in the major merger rates at roughly constant
co-moving number density are observed (Table 4),  with $\Gamma_{pairs,n_{gal}}$
and $\Re_{pairs,n_{gal}} \propto (1 + z)^{3}$.  
However, the inferred minor merger rates at constant number density 
evolve weakly with redshift. 

(6) The major-merger rates and their evolution with redshift 
for stellar-mass selected close pairs are in excellent agreement with a number of
recent galaxy evolution models (Hopkins et al. 2010b,  S08, C06, St09).    
The `total' merger rates for stellar-mass selected $G-M_{20}$ mergers is significantly higher
than the 1:1 -1:10  and 1:-1:30 stellar-mass ratio merger rates for the same models. 
It is possible that the $G-M_{20}$ merger rate is over-estimated,  or that the models
incompletely sample low-mass satellite galaxies and minor mergers.     Additional work on both the
observational and theoretical fronts is needed to resolve this issue. 

We thank L. Lin, L. de Ravel, J. Kartaltepe, and Y.
Shi for sending us detailed versions of their data and
calculations.  We thank N. Scoville,  K. Bundy, P. Hopkins, 
and the anonymous referee for helpful comments on earlier
version of this manuscript. We acknowledge the use of S. Salim's
measurements of the stellar masses for galaxies in the
Extended Groth Strip. JML acknowledges support from
the NOAO Leo Goldberg Fellowship. PJ was supported by a grant
from the W.M. Keck Foundation.
DC acknowledges receipt of a QEII Fellowship by the Australian Government.
This research used computational resources of the
NASA Advanced Supercomputing Division (NAS) and
the National Energy Research Scientific Computing Center (NERSC), which is supported by the Office of Science
of the U.S. Department of Energy.

\end{document}